\documentclass[11pt]{article}

\usepackage{amsmath,amsfonts}
\usepackage{authblk}
\usepackage{booktabs}
\usepackage{cancel}
\usepackage{commath}
\usepackage{caption}
\usepackage{float}
\usepackage[margin=1in]{geometry}
\usepackage{graphicx}
\usepackage{natbib}
\usepackage{nicefrac}
\usepackage{pdflscape}
\usepackage[section]{placeins}
\usepackage{rotating}
\usepackage{setspace}
\usepackage[usestackEOL]{stackengine}
\usepackage{subcaption}
\usepackage{url}
\usepackage[normalem]{ulem}
\usepackage{multirow}

\usepackage{pifont}
\newcommand{\xmark}{\ding{53}}%

\newcommand\vbar{\kern1pt\rule[-\dp\strutbox]{.8pt}{\baselineskip}\kern1pt}

\newcounter{algorithmtablecounter}
\newenvironment{algorithmtable}
  {%
   \refstepcounter{algorithmtablecounter}%
   \begin{table}[htb]\centering}
  {\end{table}}

\newsavebox{\largestimage}

\usepackage{tikz}
\usetikzlibrary{
	arrows.meta,
	calc,
	datavisualization.formats.functions,
	decorations.text,
	plotmarks,
	decorations.markings,
	decorations.pathreplacing,
        shapes,
        trees}
\tikzset{ shorten <>/.style={ shorten >=#1, shorten <=#1 } }

\usepackage{xcolor}
\PassOptionsToPackage{gray}{xcolor}

\newcommand{\bbeta}{\boldsymbol{\beta}}
\newcommand{\bgamma}{\boldsymbol{\gamma}}
\newcommand{\btheta}{\boldsymbol{\theta}}
\newcommand{\bbetatil}{\tilde{\boldsymbol{\beta}}}
\newcommand{\bgammatil}{\tilde{\boldsymbol{\gamma}}}
\newcommand{\bthetatil}{\tilde{\boldsymbol{\theta}}}
\newcommand{\bthetahat}{\hat{\boldsymbol{\theta}}}
\newcommand{\bthetaprime}{\boldsymbol{\theta}^\prime}

\newcommand{\bLambda}{\boldsymbol{\Lambda}}

\newcommand{\sigmahat}{\hat{\sigma}}

\newcommand{\rmd}{\mathrm{d}}

\newcommand{\bI}{\mathbf{I}}

\newcommand{\bJ}{\mathbf{J}}
\newcommand{\bP}{\mathbf{P}}

\newcommand{\bW}{\mathbf{W}}

\newcommand{\bX}{\mathbf{X}}

\newcommand{\bY}{\mathbf{Y}}

\newcommand{\bZ}{\mathbf{Z}}

\newcommand{\Qanon}{\tilde{Q}}
\newcommand{\Ztil}{\tilde{Z}}

\newcommand{\E}{\mathrm{E}}
\newcommand{\mcH}{\mathcal{H}}
\newcommand{\mcS}{\mathcal{S}}

\newcommand{\mr}[1]{\mathrm{#1}}

\newcommand{\ind}[1]{\mathbb{I}\left\{#1\right\}}
\newcommand{\Var}{\textsf{Var}}

\begin{document}

\title{Assessing treatment efficacy for interval-censored endpoints using multistate semi-Markov models fit to multiple data streams}

\author[1]{Rapha\"{e}l Morsomme}
\author[1]{C. Jason Liang}
\author[2]{Allyson Mateja}
\author[1]{Dean A. Follmann}
\author[3]{Meagan P. O'Brien}
\author[3]{Chenguang Wang}
\author[4]{Jonathan Fintzi\footnote{Corresponding author; jonathan.fintzi@bms.com}}

\affil[1]{Biostatistics Research Branch, National Institute of Allergy and Infectious Diseases, National Institutes of Health, Bethesda, MD, U.S.A.}
\affil[2]{Clinical Monitoring Research Program Directorate, Frederick National Laboratory for Cancer Research, Frederick, MD, U.S.A.}
\affil[3]{Regeneron Pharmaceuticles, Tarrytown, NY, U.S.A.}
\affil[4]{Statistical Methodology and Innovation, Bristol Myers Squibb, Lawrenceville, NJ, U.S.A.}

\maketitle


\begin{abstract}
We introduce a computationally efficient and general approach for utilizing multiple, possibly interval-censored, data streams to study complex biomedical endpoints using multistate semi-Markov models. Our motivating application is the REGEN-2069 trial, which investigated the protective efficacy (PE) of the monoclonal antibody combination REGEN-COV against SARS-CoV-2 when administered prophylactically to individuals in households at high risk of secondary transmission. Using data on symptom onset, episodic RT-qPCR sampling, and serological testing, we estimate the PE of REGEN-COV for asymptomatic infection, its effect on seroconversion following infection, and the duration of viral shedding. We find that REGEN-COV reduced the risk of asymptomatic infection and the duration of viral shedding, and led to lower rates of seroconversion among asymptomatically infected participants. Our algorithm for fitting semi-Markov models to interval-censored data employs a Monte Carlo expectation maximization (MCEM) algorithm combined with importance sampling to efficiently address the intractability of the marginal likelihood when data are intermittently observed. Our algorithm provide substantial computational improvements over existing methods and allows us to fit semi-parametric models despite complex coarsening of the data.
\end{abstract}


\section{Introduction}
\label{sec:introduction}
The primary objective of late-phase vaccine and therapeutic trials is to determine if an intervention reduces the risk of objective clinical endpoints, symptomatic disease, or death \citep{mcleod2019choosing}. Secondary analyses examine how the intervention alters the dynamics of clinical progression and assess the treatment effect on mechanistic endpoints \citep{fdaguidance2017multiple}. The REGEN-2069 trial found that prophylactic treatment with REGEN-COV, a combination of the monoclonal antibodies (mAb) casirivimab and imdevimab, reduced the risk of symptomatic SARS-CoV-2 infection among individuals at high risk of exposure \citep{obrien2021regen}. Our goals are to assess the PE of REGEN-COV against infection, including asymptomatic infection, to characterize its effect on seroconversion, and estimate the duration of detectable viral shedding. 

In REGEN-2069, the day of symptom onset was observed and SARS-CoV-2 infection was confirmed by nasopharyngeal swab for reverse-transcriptase–quantitative polymerase-chain-reaction testing (PCR). Viral shedding was assessed weekly via PCR to detect ongoing, possibly asymptomatic, infection. Though episodic PCR testing captures the prevalence of ongoing infection at scheduled times, it may not capture infections with short shedding durations. Serological testing was conducted at baseline and the end of the efficacy assessment period (EAP) for anti-nucleocapsid IgG antibodies.  In this setting, seroconversion refers to the detection of such antibodies in previously na\"{\i}ve participants and reflects an immune response to infection, possibly asymptomatic. Whether an infected participant seroconverts may depend on their symptom and vaccination history; not all infected individuals seroconvert \citep{follmann2022antinucleocapsid}. Hence, the timing of infection is temporally coarse and its extent incompletely captured by symptom onset, PCR, or serology alone. Our goal in assimilating multiple data streams is to estimate the effect of mAb on difficult to measure endpoints implicated by asymptomatic infection and on the timing and duration of shedding of detectable virus; these parameters are not estimable without unifying multiple data sources to better capture clinical and immunological progression among participants.

Multistate models characterize the time-evolution of a biological process through discrete biological states and are natural tools for data assimilation since the model states may be informed by multiple data sources. Observations of continuous observed processes are amenable to analysis with standard parametric and non-parametric methods \citep{andersen2002multi,cook2018multistate}. Multistate likelihoods for intermittently observed processes involve computationally intractable transition probabilities except in the case of Markov processes where transition probabilities can be obtained by solving the Kolmogorov forward equations via a matrix exponential \citep{kalbfleisch1985analysis} or by numerical integration \citep{titman2011flexible}. A semi-Markov assumption allows transition intensities to depend on the duration of state occupancy and other aspects of the history up to the current state. Misspecification of a model's operational time scale, e.g., by assuming that a model is time-inhomogeneous Markov rather than semi-Markov, can bias estimates of key functionals, such as sojourn time distributions \citep{cheung2022multistate}. Unfortunately, transition probabilities can no longer be easily computed for semi-Markov models. One strategy is to approximate a semi-Markov process with a latent continuous-time Markov chain (CTMC) with Coxian phase-type structure that induces occupancy-time dependent sojourns \citep{titman2010semi,lange2015joint}. Phase-type latent CTMCs may not be  identifiabile when the true process is nearly Markovian or the data cannot resolve more complicated temporal dynamics \citep{titman2010semi}. Existing sampling-based ``data augmentation" approaches numerically integrate over the missing paths \citep{aralis2019stochastic,barone2022bayesian}, but become computationally expensive and fragile for complex models and when the fraction of missing information is large \citep{papaspiliopoulos2007general}.  

We develop a computationally efficient Monte Carlo expectation maximization (MCEM) framework for fitting multistate semi-Markov models when the data are coarsened or subject to measurement error. Our framework is general, rather than model-specific, and explicitly allows for incomplete observation of the state at assessment times and for data to be accrued under continuous and intermittent observation schemes. We incorporate covariate effects and spline-based transition intensities for semi-parametric inference. Our methods allow for a novel and detailed study of the effect of REGEN-COV on endpoints related to asymptomatic infection, leading to a more complete understanding of its mechanism and clinical impact.

\section{Data and Modeling framework}
\label{sec:modelingframework}

\subsection{Motivating example: REGEN-COV mAb for COVID-19 prophylaxis} 
\label{subsec:regen2069descrip}
REGEN-2069 was a randomized placebo-controlled trial to evaluate the REGEN-COV mAb as prophylaxis against symptomatic SARS-CoV-2 infection in unvaccinated persons at high risk of infection from a confirmed case within their household \citep{obrien2021regen}. Participants, aged 12 and older with no prior SARS-CoV-2 infection by serology and PCR, were enrolled within 96 hours of a household member's diagnosis of SARS-CoV-2 infection. Participants were monitored for symptoms and SARS-CoV-2 infection was confirmed by PCR at presentation. Participants' infection status was also assessed weekly by PCR for viral shedding, and serological testing conducted at the end of the EAP to identify seroconversion following infection.

\begin{figure}[htbp]
    \begin{subfigure}[b]{\textwidth}
    \centering\small
    \begin{tabular}{cccccclcc}
        \toprule
        &&& \multicolumn{3}{c}{Observed data at end time} & & \multicolumn{2}{c}{Model state}\\
        \cmidrule(lr){4-6} \cmidrule(lr){8-9} \shortstack{ID} & \shortstack{Start\\time} & \shortstack{End\\time} & PCR+ & \shortstack{Ever \\ sym.} & Sero.+ & Derived information & \shortstack{State\\from} & \shortstack{State\\to}\\
         \cmidrule(lr){1-9} 001 & 0 & 7 & + & -- & $\cdot$ & Infected in (0, 7] & 1 & 2 \\
         001 & 7 & 14 & + & -- & $\cdot$ & Still PCR+  & 2 & 2\\
         001 & 14 & 21 & -- & -- & $\cdot$ & PCR+ to PCR$-$ in $3^\mr{rd}$ week & 2 & 3\\
         001 & 21 & 28 & -- & -- & -- & No seroconversion & 3 & 3 \\
         \cmidrule(lr){1-9} 002 & 0 & 7 & -- & -- & $\cdot$ & No infection in (0, 7] & 1 & 1 \\
         002 & 7 & 9-$\epsilon$ & + & + & $\cdot$ & Infected in (7, 9-$\epsilon$] & 1 & 2 \\
         002 & 9-$\epsilon$ & 9 & + & + & $\cdot$ & Symptom onset on D9 & 2 & 4 \\
         002 & 9 & 14 & -- & + & $\cdot$ & PCR+ $\rightarrow$ PCR$-$ in (9, 14] & 4 & 4\\
         002 & 14 & 21 & -- & + & $\cdot$ & No change in PCR-- & 4 & 5\\
         002 & 21 & 28 & -- & + & -- & No seroconversion & 5 & 5\\
         \cmidrule(lr){1-9} 003 & 0 & 7 & -- & -- & $\cdot$ & Na\"{\i}ve at D7 or PCR+ $\rightarrow -$ in (0,7] & 1 & \{1,3\}\\
         003 & 7 & 14 & -- & -- & $\cdot$ &  Na\"{\i}ve at D14 or PCR+ $\rightarrow -$ in (7,14] & \{1,3\} & \{1,3\}\\
         003 & 14 & 21 & -- & -- & $\cdot$ & Na\"{\i}ve at D21 or PCR+ $\rightarrow -$ in (14, 21] & \{1,3\} & \{1,3\}\\
         003 & 21 & 28 & -- & -- & + & \Centerstack[l]{Seroconversion at D28 $\implies$ prior infec. \\ and short shedding duration} & \{1,3\} & 3\\
        \bottomrule
    \end{tabular}
    \captionsetup{skip=0.5em}
    \caption{Combining PCR assessment, symptom history, and serology data to deduce a patient's infection history.}
    \label{tab:exampledata_info}
    \end{subfigure} \\\vfill
    
    \begin{subfigure}[b]{0.5\textwidth}
    \centering\small
    \begin{tabular}{l}
        \toprule
         Modeling assymptions \\
         \cmidrule(lr){1-1} 
            1. Participants who never become symptomatic,\\\hspace{0.9em} PCR+, or seropositive are uninfected. \\
            2. PCR+ precedes symptoms precede seroconversion. \\
            3. Symptom onset precedes PCR+ $\longrightarrow$ PCR$-$. \\
            4. No transitions from PCR$-$ $\longrightarrow$ PCR+. \\
            5. No sero-reversion on study.  \\
         \bottomrule
    \end{tabular}
    \captionsetup{skip=0.5em}
    \caption{Assumptions used to identify infections and derive possible transitions between states of infection, symptom onset, and seroconversion.}
    \label{tab:regenassump}
    \end{subfigure}
    \hfill
    \begin{subfigure}[b]{0.45\textwidth}
        \centering
        \begin{tikzpicture}[scale=0.95, every node/.style={scale=0.9}, every text node part/.style={align=center}]
        \definecolor{cbgray1}{gray}{0.5}
        \definecolor{cbgray2}{gray}{0.25}

        \draw (-1, 0.0) node[align=center, draw,rectangle,thick] (1) {1};
        \draw (0.75, 0.0) node[align=center, draw, rectangle] (2) {2};
        \draw (3.25, 0.0) node[align=center, draw, rectangle] (3) {3};
        \draw (0.75, -2) node[align=center, draw, rectangle] (4) {4};
        \draw (3.25, -2) node[align=center, draw, rectangle] (5) {5};
        \draw (5.5, 0.0) node[align=center, draw, ellipse, dashed] (6) {N--};
        \draw (5.5, -2) node[align=center, draw, ellipse, dashed] (7) {N+};

        \draw [thick,->,shorten >=3mm,shorten <=1mm] (1) to (2) node[midway,above] {};
        \draw [thick,->,shorten >=1.5mm,shorten <=1.5mm] (2) to (3) node[midway,above] {};
        \draw [thick,->,shorten >=2mm,shorten <=1.5mm] (2) to (4) node[midway,above] {};
        \draw [thick,->,shorten >=1.5mm,shorten <=1.5mm] (4) to (5) node[midway,above] {};
        
        \draw [thick,-,shorten <=3mm,dashdotted] (3.east) to (7.north west) node {};
        \draw [thick,-,shorten <=2mm,dashdotted] (5.east) to (7.west) node {};
        \draw [thick,-,shorten <=2mm,dashdotted] (3.east) to (6.west) node {};
        \draw [thick,-,shorten <=3mm,dashdotted] (5.east) to (6.south west) node {};

        \draw[semithick, densely dotted] (0.315, -0.35) rectangle (3.65, 0.425);
        \draw[semithick, densely dotted] (0.315,-2.45) rectangle (3.65, -1.65);

        \draw[very thin, densely dashed] (0.425,-2.35) rectangle (1.075, 0.325);
        \draw[very thin, densely dashed] (2.92,-2.35) rectangle (3.575, 0.325);
            
        \node[anchor=north] at (-1,-0.4) {Na\"{\i}ve};
        \node[anchor=north] at (2, -0.35) {No Sympt.};
        \node[anchor=north] at (2, -2.45) {Ever Sympt.};
        \node[anchor=south] at (0.9,0.4) {PCR+};
        \node[anchor=south] at (3.4,0.4) {PCR$-$};
    \end{tikzpicture}
    \captionsetup{skip=0.5em}
    \caption{Model for infection, symptoms, and seroconversion. Solid arrows represent transitions. Dash-dotted lines depict final serology status.}
    \label{subfig:regendiagram}    
    \end{subfigure}
    \caption{(\ref{tab:exampledata_info}) Raw data and derived model states for 3 hypothetical study participants. (\ref{tab:regenassump}) Modeling assumptions. (\ref{subfig:regendiagram}) Multistate model transition diagram.}
    \label{fig:regenfigures}
\end{figure}
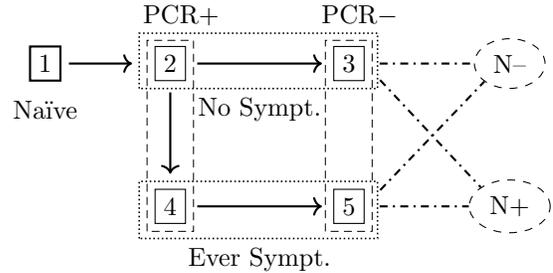

The data accrue at discrete times and are a coarsened reflection of the natural history of infection and subsequent clinical and immune response. Table~\ref{tab:exampledata_info} illustrates how PCR assessments, symptom history, and serology data combine to enrich the data for three hypothetical participants. Participant 001 was PCR+ on day 7, and hence infected in the first week, but did not manifest symptoms or seroconvert. Participant 002 was symptomatic and had infection confirmed by PCR at symptom onset, though their infection would not have been detected on the basis of regularly scheduled PCR assessments and final serology. Combining their PCR and symptom history narrows their infection window to day 7, when they were PCR$-$, to day 9, when their symptoms manifested. Their symptom onset is encoded as occurring in the interval, (9-$\epsilon$, 9] where $\epsilon$ is an arbitrary duration that is short relative to the time-resolution of the data, e.g., 1 hour. Participant 002 stopped shedding detectable virus by day 14. Participant 003 was confirmed asymptomatically infected by final serology. We do not know whether they were infection na\"{\i}ve at their interim assessments since individuals with short shedding durations might escape detection by weekly PCR.

Table~\ref{tab:regenassump} summarizes the assumptions under which we formulate a biological model for infection, symptoms, and immunological response. Evidence of SARS-CoV-2 infection may be captured by positive PCR, manifestation of symptoms, or positive serology. Participants who never manifest symptoms or test positive are assumed to be infection na\"{\i}ve at the end of the EAP, i.e., unaffected implies never infected. In this spirit, we treat infection as the occurrence of a participant being measurably affected, as evidenced by viral shedding detectable by nasopharyngeal RT-qPCR. Thus, in our model, infection is the transition from na\"{\i}ve to PCR+. The other modeling assumptions reflect our understanding of the temporal precedence of viral shedding, symptoms, and seroconversion. For instance, SARS-CoV-2 infected individuals shed detectable virus prior to manifesting symptoms \citep{puhach2023sars}. These assumptions inform the formulation of the model states and allowable transitions depicted in Figure~\ref{subfig:regendiagram}, and identify the model state, or set of possible model states, at each observation time, as shown in the last two columns of Table~\ref{tab:exampledata_info}. 

Before proceeding, we highlight several aspects of the data that complicate statistical inference. First, the data intermingle continuous observations of one aspect of the process, symptom onset, with intermittent observations of other aspects of progression, namely, infection and viral shedding. Second, the true state for participants whose infection is confirmed by serology alone is unknown at interim assessment times. Finally, the data do not preclude the occurrence of multiple transitions within an inter-observation interval and there is some variability in assessment times around scheduled assessments, e.g., a PCR assessment scheduled for study day 7 could have been conducted at day 9. In consideration of these complications, the next sections will develop a framework using continuous-time models for the latent biological process, though we note a similar development is possible for multistate processes that evolve discretely in time.

\subsection{Data for continuously and intermittently observed multistate processes}
\label{subsec:data}

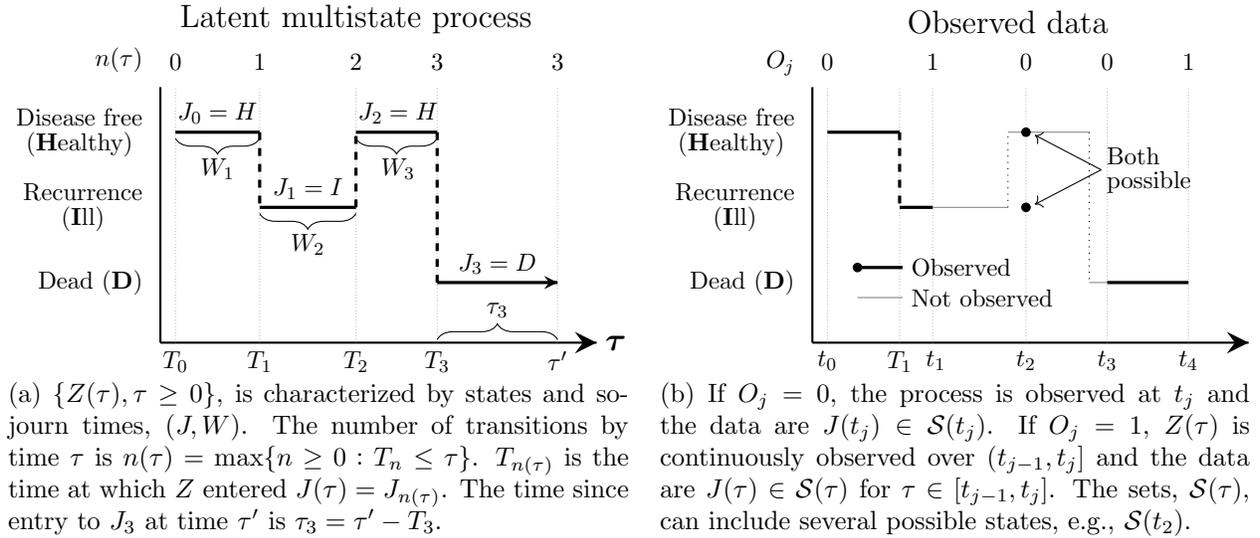
\begin{figure}[htbp]
\centering
\setkeys{Gin}{width=\linewidth}
    \begin{subfigure}[t]{0.5\textwidth}
        \begin{tikzpicture}[scale=0.8,every node/.style={scale=0.85}]
        \definecolor{cbgray1}{gray}{0.5}
        \definecolor{cbgray2}{gray}{0.25}

        \draw (3, 3.75) node[anchor=south] {\Large Latent multistate process};

        \draw [thick] (-0.25,-1.25) -- (-0.25,3);
        \draw[decoration={markings,mark=at position 1 with {\arrow[scale=2,>=stealth]{>}}},postaction={decorate},thick] (-0.25,-1.25) -- (7,-1.25);
        \draw (7, -1.25) node[anchor=west] {\Large$\boldsymbol{\tau}$};

        \foreach \i in {0.0, 2.8, 4.5, 6.35} \draw [very thin, densely dotted, color = cbgray1] (\i,-1.25)--(\i,3);
        
        \draw (0.0, -1.25) node[anchor = north] {$T_0$};
        \draw (2.8, -1.25) node[anchor = north] {$T_1$};
        \draw (4.5, -1.25) node[anchor = north] {$T_2$};
        \draw (6.35, -1.25) node[anchor = north] {$\tau^\prime$};

        \draw (-0.4, 2.25) node[anchor = east] {\shortstack{Disease free\\ (\textbf{H}ealthy)}};
        \draw (-0.4, 1) node[anchor = east] {\shortstack{Recurrence\\(\textbf{I}ll)}};
        \draw (-0.4, -0.25) node[anchor = east] {Dead (\textbf{D})};

        \draw [very thick] (0.0, 2.25) -- (2.8, 2.25);
        \draw [very thick] (2.8, 1) -- (4.5, 1);
        \draw [very thick, decoration={markings,mark=at position 1 with {\arrow[scale=1,>=stealth]{>}}},postaction={decorate}] (4.5, -0.25) -- (6.35, -0.25);

        \draw [very thick, dashed] (2.8, 2.25) -- (2.8, 1);
        \draw [very thick, dashed] (4.5, 1) -- (4.5, -0.25);

        \draw (1.4, 2.25) node[anchor = south] {$J_0 = H$};
        \draw (1.4, 1.95) node[anchor = north] {$W_1$};
        \draw [decorate,decoration={brace,amplitude=6pt,mirror}] (0.0, 2.15) -- (2.8, 2.15);

        \draw (3.65, 1) node[anchor = south] {$J_1 = I$};
        \draw (3.65, 0.7) node[anchor = north] {$W_2$};
        \draw [decorate,decoration={brace,amplitude=6pt,mirror}] (2.8,0.9) -- (4.5,0.9);

        \draw (5.45, -0.25) node[anchor = south] {$J_2 = D$};
        \draw (5.45, -0.45) node[anchor = north] {$\tau_{2}$};
        \draw [decorate,decoration={brace,amplitude=6pt}] (4.5,-1.2) -- (6.35,-1.2);

        \draw (-0.4, 3.45) node[anchor = east] {$n(\tau)$};
        \draw (0, 3.15) node[anchor = south] {0};
        \draw (2.8, 3.15) node[anchor = south] {1};
        \draw (4.5, 3.15) node[anchor = south] {2};
        \draw (6.35, 3.15) node[anchor = south] {2};
    \end{tikzpicture}
    \caption{$\{Z(\tau),\tau\geq 0\}$, is characterized by states and sojourn times, $(J,W)$. The number of transitions by time $\tau$ is $n(\tau) = \max\{n\geq 0 : T_n \leq \tau\}$. $T_{n(\tau)}$ is the time at which $Z$ entered $J(\tau) = J_{n(\tau)}$. The time since entry to $J_2$ at time $\tau^\prime$ is $\tau_2 = \tau^\prime - T_2$.}
    \label{fig:processnotation}
    \end{subfigure}
    \hfill
    \begin{subfigure}[t]{0.475\textwidth}
        \begin{tikzpicture}[scale=0.8,every node/.style={scale=0.85}]
        \definecolor{cbgray1}{gray}{0.5}
        \definecolor{cbgray2}{gray}{0.25}

        \draw (3, 3.75) node[anchor=south] {\Large Observed data};

        \draw [thick] (-0.25,-1.25) -- (-0.25,3);
        \draw[decoration={markings,mark=at position 1 with {\arrow[scale=2,>=stealth]{>}}},postaction={decorate},thick] (-0.25,-1.25) -- (7,-1.25);

        \foreach \i in {0.0, 2, 3.75, 4.5, 6} \draw [very thin, densely dotted, color = cbgray1] (\i,-1.25)--(\i,3);
        \draw (0.0, -1.25) node[anchor = north] {$t_0$};
        \draw (2.8, -1.25) node[anchor = north] {$T_1$};
        \draw (2, -1.25) node[anchor = north] {$t_1$};
        \draw (3.75, -1.25) node[anchor = north] {$t_2$};
        \draw (4.5, -1.25) node[anchor = north] {$t_3$};
        \draw (6, -1.25) node[anchor = north] {$t_4$};

        \draw (-0.4, 2.25) node[anchor = east] {\shortstack{Disease free\\ (\textbf{H}ealthy)}};
        \draw (-0.4, 1) node[anchor = east] {\shortstack{Recurrence\\(\textbf{I}ll)}};
        \draw (-0.4, -0.25) node[anchor = east] {Dead (\textbf{D})};



        \draw [very thick] (0.0, 2.25) -- (2, 2.25);
        \draw [very thin, color=cbgray1] (2, 2.25) -- (2.8, 2.25);
        \draw [very thin, color=cbgray1] (2.8, 1) -- (4.5, 1);
        \draw [very thick, decoration={markings,mark=at position 1 with {\arrow[scale=1,>=stealth]{>}}},postaction={decorate}] (4.5, -0.25) -- (6.35, -0.25);

        \draw [thin, dotted] (2.8, 2.25) -- (2.8, 1);
        \draw [thin, dotted] (4.5, 1) -- (4.5, -0.25);

        \filldraw[black] (3.75, 2.25) circle (2pt) node {};
        \filldraw[black] (3.75, 1) circle (2pt) node {};

        \draw (-0.4, 3.4) node[anchor = east] {$O_j$};
        \draw (0, 3.15) node[anchor = south] {0};
        \draw (2, 3.15) node[anchor = south] {1};
        \draw (3.75, 3.15) node[anchor = south] {0};
        \draw (4.5, 3.15) node[anchor = south] {0};
        \draw (6, 3.15) node[anchor = south] {1};

        \draw [thin, decoration={markings,mark=at position 1 with {\arrow[scale=1]{>}}},postaction={decorate}] (4.55, 1.625) -- (3.9, 1.1);
        \draw [thin, decoration={markings,mark=at position 1 with {\arrow[scale=1]{>}}},postaction={decorate}] (4.55, 1.625) -- (3.9, 2.15);
        \draw (4.5, 1.625) node [anchor=west,align=left] {\shortstack[l]{Both\\possible}};

        \filldraw (0.5, 0) circle (2pt) node {};
        \draw [very thick] (0.5, 0) -- (1.25, 0); 
        \draw (1.25, 0) node [anchor=west] {Observed};
        
        \draw [thin, color=cbgray1] (0.5, -0.5) -- (1.25, -0.5); 
        \draw (1.25, -0.5) node [anchor=west] {Not observed};
        
        \end{tikzpicture}
        \caption{If $O_j = 0$, the process is observed at $t_j$ and the data are $J(t_{j}) \in \mcS(t_j)$. If $O_j = 1$, $Z(\tau)$ is continuously observed over $(t_{j-1},t_j]$ and the data are $J(\tau) \in \mcS(\tau)$ for $\tau\in[t_{j-1},t_j]$. The sets, $\mcS(\tau)$, can include several possible states, e.g., $\mcS(t_2)$.}
        \label{fig:datanotation}
    \end{subfigure}
    \caption{Notation for a latent multistate process (\ref{fig:processnotation}) and  data (\ref{fig:datanotation}). The true  path, $ \left\{(H,0),\ (I,W_1),\ (D,W_2)\right\}$, is continuously observed from $t_0$ until $t_1$. The participant is observed to be alive at $t_2$ and the exact time of death is recorded at $t_3$. Hence, we observe  $\left[\{J(t_0) = H\},\ \{J(\tau) = H,\ \tau \in (t_0,t_1]\}, \{J(t_2)\in(H,I)\},\ \{J(t_3) = D\}\right]$.}
    \label{fig:notation}
\end{figure}

We proceed to establish notation for representing data from a latent multistate process  under a mixture of intermittent and continuous observation schemes while allowing for incomplete observation of the true state. Figures \ref{fig:processnotation} and \ref{fig:datanotation} provide a visual reference using a progressive illness-death process where disease recurrence is incompletely observed. 

We let $ \{Z(\tau),\ \tau\geq0\} $ denote a latent multistate process that evolves continuously in time, $\tau$, and takes values in a state space, $ \mcS = \{{s_1,\dots,s_K} \}$. As diagrammed in Figure \ref{fig:processnotation}, a sample path of $ Z $ is represented as $ \{(J_n,W_n),\ n\geq 0\} $, where $ J_n\in\mcS $ is the state after $n$ transitions and $ W_{n+1} \in \mathbb{R}^+$ is the waiting time spent in $ J_{n} $ before transitioning to $ J_{n+1} $. By convention, $ W_0 = 0 $. The time at which the $ n^{th} $ transition occurs is $ T_n = \sum_{j = 1}^n W_j $. The number of transitions by time $\tau$ is denoted $ n(\tau) = \max\{n\geq 0 : T_n \leq \tau\} $. The state at time $\tau$ is $ J(\tau) = J_{n(\tau)} $, and $ \tau_n  = \tau - T_{n(\tau)} $ is the time since entry to $ J(\tau) $. The path of $Z$ over the interval $[\tau_r,\tau_s]$, is denoted $Z(\tau_r,\tau_s)$. We write $ Z_i $ or $ (J,W)_i $ for participant $i$'s sample path, and $ \bZ $, or $ (\bJ,\bW) $, to denote the paths for all participants. 

The data for participant $i$ accrue at times $t_{i,0},\dots,t_{i,L_i} $. Let $O_{i,j} = 0$ if we observe $J(t_{i,j})\in\mcS(t_{i,j})$, where $\mcS(t_{i,j}) \subseteq \mcS$ are states consistent with the data at $t_{i,j}$. Let $O_{i,j}=1$ if the $Z_i$ is continuously observed over $(t_{i,{j-1}},t_{i,j}]$. Specifically, we observe $J(\tau)\in\mcS(\tau)$ for $\tau\in(t_{i,j-1},t_{i,j}]$, which we denote using the shorthand, $J(t_{i,j-1}, t_{i,j})\in\mcS(t_{i,j-1},t_{i,j})$. In Figure \ref{fig:datanotation}, we observe that the participant is healthy from $t_0$ to $T_1$, when their disease recurs, and that they are ill during $(T_1, t_1]$. We then observe that the participant is alive at $t_2$, but not whether their disease has recurred, so $\mcS(t_2) = \{\mr{Healthy}, \mr{Ill}\}$. The participant is known to have died by $t_3$.  Let $\bY_{i,j} = \left(\mcS(t_{i,j-1}, t_{i,j}), O_{i,j}\right)$ denote the data for participant $i$ over the interval $(t_{i,j-1},t_{i,j}]$, and $\bY_i = \left(\bY_{i,j}\right)_{j=0,\dots,L_i}$. Finally, $\bY = \left(\bY_i\right)_{i=1,\dots,N}$ is the data for all study participants.

\subsection{Multistate semi-Markov models}
\label{subsec:multistate}

The time-evolution of a multistate process is governed by transition intensity functions 
\begin{align}
	\label{eqn:transition_intensity}
	\lambda_{k\ell}\left(\tau,\mcH(\tau^-), \bX(\tau^-), \btheta\right) &= \lim_{\Delta\tau \downarrow 0} \frac{\Pr\left (Z(\tau^- + \Delta\tau) = \ell \mid Z(\tau^-) = k, \mcH(\tau^-), \bX(\tau^-), \btheta\right )}{\Delta \tau}, 
\end{align}
for $ k\neq \ell $, where $ \mcH(\tau^-) $ is the history of the process up to $ \tau $, $\bX(\tau^-)$ are covariates whose values are known at $\tau^-$, and $ \btheta = (\bgamma,\bbeta) $ is a vector of parameters that we partition into those that relate to the baseline transition intensities and covariates, respectively. 

A convenient parameterization assumes that covariates act multiplicatively upon the intensity, which under a semi-Markov assumption depends on the time since state entry, $ \tau_n $. Hence, 
\begin{align}
    \label{eqn:semimarkovintensity}
    \lambda_{k\ell}\left(\tau, \mcH(\tau^-), \bX(\tau^-),\btheta\right) = \lambda_{k\ell}\left(\tau_n, \bX(\tau^-),\btheta\right) = \eta_{k\ell}\left(\tau_n, \bgamma\right)\xi_{k\ell}\left(\bX(\tau^-),\bbeta\right), 
\end{align} 
where $\eta_{k\ell}\left(\tau_n,\bgamma\right)\in\mathbb{R}^+$ depends on the time since entry to the  state at $\tau$, and $\xi_{k\ell}\left(\bX(\tau^-),\bbeta\right)\in\mathbb{R}^+$ captures the effect of covariates, possibly including study time. 

The hazard of exiting state $k$ after a sojourn of length $\tau_n$ is the sum of intensities to states reachable from $k$, i.e., $ \lambda_{k}\left(\tau_n, \bX(\tau^-),\btheta\right) = \sum_{\ell \neq k}\lambda_{k\ell}\left(\tau_n, \bX(\tau^-),\btheta\right) $. The probability of remaining in the state for time $\tau_n$ since entry is
\begin{align}
	\label{eqn:survival}
	S_k(\tau_n\mid\bX,\btheta) = \exp\left (-\int_{0}^{\tau_n}\lambda_{k}\left(u, \bX(u^-),\btheta\right)\rmd u\right ).
\end{align}

Let $N_i$ be the number of states visited by participant $i$ over an observation period, $ [0,C_i] $, where $ C_i = \min(A, R_i)$ is the time of administrative censoring, $A$, or random censoring, $R_i$. $T_{i,N_i}$ is the minimum of the time at which participant $i$ enters an absorbing state or is censored, and $W_{i,N_i}$ is the time spent in the last observed state up to $T_{N_i}$. Set $\delta_{i,n} = 1$ if a transition occurs at time $T_{i,n}$ and 0 otherwise. The probability density of a continuously observed sample path is a product of intensities at transition times and waiting time probabilities \citep{cook2018multistate}:
\begin{align}
    f_i\left(Z_i\mid \bX_i,\btheta\right) &= \Pr(J_{i,0}\mid \bX_i, \btheta)\prod_{n=1}^{N_i} \Pr((J_{i,n},W_{i,n})\mid \bX_i,\btheta),\nonumber\\
    & = \Pr(J_{i,0}\mid \bX_i, \btheta)\prod_{n = 1}^{N_i} \lambda_{J_{i,n-1}J_{i,n}}\left(W_{i,n}, \bX_i(T_{i,n}^-),\btheta\right)^{\delta_{i,n}} \times S_{J_{i,n-1}}(W_{i,n}\mid \bX_i,\btheta).
    \label{eqn:exactlikelihood}
\end{align}
For $ P $ independent participants, $f(\bZ\mid\bX,\btheta) = \prod_{i=1}^P f_i(Z_i\mid\bX_i, \btheta)$.

Suppressing the dependence on $\bX$ and $\btheta$, the marginal probability density for an individual when observation is either intermittent or continuous is
\begin{align}
    g_i(\bY_i) &= \Pr\left(J_{i,0}\in\mcS(t_{i,0})\right)\prod_{j=1}^{L_i} \left[\Pr\left(J(t_{i,j})\in\mcS(t_{i,j})\mid\bY_{i,0:(j-1)}\right)^{(1 - O_{i,j})} \times \right.\nonumber\\ 
    &\hspace{12em} \left. \Pr\left(J(t_{i,j-1},t_{i,j})\in\mcS(t_{i,j-1},t_{i,j})\mid \bY_{i,0:(j-1)}\right)^{O_{i,j}}\right],\label{eqn:datalikelihoodprobs}\\
    &= \int\ind{J_{i,0}\in\mcS(t_{i,0})}\prod_{j=1}^{L_i} \left[\ind{J(t_{i,j})\in\mcS(t_{i,j})}^{(1 - O_{i,j})} \times \right.\nonumber\\ 
    &\hspace{12em} \left. \ind{J(t_{i,j-1},t_{i,j})\in\mcS(t_{i,j-1},t_{i,j})}^{O_{i,j}}\right]f_i(Z)\ \rmd Z,\label{eqn:datalikelihood}
\end{align}
which is the expectation of indicators for whether a sample path is concordant with the data taken over the distribution of $Z$. For $P$ independent participants, $g(\bY) = \prod_{i=1}^P g_i(\bY_i)$. 

\subsection{Maximum likelihood via Monte Carlo expectation-maximization}
\label{subsec:mcem}

Direct optimization of the marginal likelihood is challenging when data are intermittently observed since the number and timing of transitions are unknown. The expectation-maximization (EM) algorithm of \citet{dempster1977maximum} facilitates maximum likelihood estimation in missing data settings by alternating between an E-step that calculates the expected complete data log-likelihood, referred to as the Q-function, and an M-step that maximizes the Q-function. Let $\bthetahat^{(t-1)}$ be the estimate of $\btheta$ after $t-1$ steps of the EM algorithm. The Q-function is,
\begin{align}
    \label{eqn:q}
    Q\left(\btheta,\bthetahat^{(t-1)}\right) &= \mr{E}_{\bZ\mid \bY, \bX, \bthetahat^{(t-1)}}\left[\ell\left(\bZ, \bY\mid \bX,\btheta\right)\right],
\end{align}
where $\ell\left(\bZ, \bY\mid \bX,\btheta\right)=\log(f(\bZ\mid\bX,\btheta)p(\bY\mid \bZ, \bX, \btheta))$, with $p(\bY\mid \bZ, \bX, \btheta)=\ind{\bZ \textsf{ concords with} \bY}$.

For multistate semi-Markov models, the Q-function does not admit a closed-form expression. The Monte Carlo expectation-maximization (MCEM) algorithm, replaces the E-step of the EM algorithm with a Monte Carlo estimate of the Q-function \citep{wei1990monte}. For each participant, $i=1,\dots,P$, we sample paths, $Z_i^{(1)},\dots,Z_i^{(M_i)}$, from a proposal distribution with density $h_i^\prime\left(Z_i^{(m)}\mid \bX_i,\bthetaprime,\bY_i\right)$. The importance sampling estimate of (\ref{eqn:q}) in the $t^{\mathrm{th}}$ E-step is
\begin{align}
    \label{eqn:qfun}
    \Qanon\left(\btheta,\bthetahat^{(t-1)}\right) &= \sum_{i=1}^P\frac{\sum_{m=1}^{M_i} \nu_{i,m}^{(t-1)}l_i\left(Z_i^{(m)}, \bY_i\mid \bX_i,\btheta\right)}{\sum_{j=1}^{M_i} \nu_{i,j}^{(t-1)}},
\end{align}
where the self-normalized importance weights simplify to
\begin{align}
    \label{eqn:importance-weight}
    \nu_{i,m}^{(t-1)} 
    = \frac{f_i\left(Z_i^{(m)}\mid \bX_i,\bthetahat^{(t-1)}\right)}{h_i\left(Z_i^{(m)} \mid \bX_i,\bthetaprime\right)},
\end{align}
as shown in Section \ref{subsec:importanceweights} of the Supplementary Material. The complete data log-likelihood evaluates to the log-likelihood of a continuously observed sample path,
\begin{align*}
    l_i\left(Z_i^{(m)}, \bY_i\mid \bX_i,\btheta\right)
    &= \log \left(f_i\left(Z_i^{(m)}\mid \bX_i,\bthetahat^{(t-1)}\right) p_i(\bY_i\mid Z_i^{(m)}, \bX_i, \btheta)\right), \nonumber \\
    &= \log \left(f_i\left(Z_i^{(m)}\mid \bX_i,\bthetahat^{(t-1)}\right) \ind{Z_i^{(m)} \textsf{ concords with }\bY_i} \right), \nonumber \\
    &= \log \left(f_i\left(Z_i^{(m)}\mid \bX_i,\bthetahat^{(t-1)}\right) \right),    
\end{align*}
since $Z_i^{(m)}$ is sampled conditionally on the data (Section~\ref{subsec:proposal_framework}). In practice, we use Pareto smoothing \citep{vehtari2015pareto} to shrink extreme importance weights, which may be unstable for certain parametric baseline intensities.

The ascent property of the EM algorithm that assures $L\left(\bY\mid\bthetahat^{(t)}\right)\geq L\left(\bY\mid\bthetahat^{(t-1)}\right)$  is no longer guaranteed in MCEM. We implement the ascent MCEM algorithm of \citet{caffo2005ascent}, which recovers the ascent property with high probability by accepting the maximizer in each M-step only when the increase in the Q-function is distinguished from Monte Carlo error with sufficient confidence. The ascent MCEM algorithm calls to augment the pool of Monte Carlo samples when the change in the Q-function is statistically indistinguishable from zero. The Monte Carlo standard of the change in the Q-function is a function of the effective sample size (ESS), 
\begin{equation} 
    \label{eqn:ess}
    M^\star_i = \frac{1}{\sum_{m=1}^{M_i} \bar{\nu}_{i,m}^2},
\end{equation}
where $\bar{\nu}_{i,m} = \nu_{i,m} / \sum_{j=1}^{M_i}\nu_{i,j}$ are the self-normalized importance weights \citep{liu2001monte}. Early iterations of the MCEM algorithm will typically yield large increases in the Q-function. Hence, setting the initial ESS at some modest target, e.g., $M_i^\star = 10$ to $25$, avoids wasteful computation when the algorithm is far from the MLE. We suggest a larger starting ESS, e.g., $M_i^\star =50$ to $100$, for complicated models to help the algorithm avoid local modes. The ascent MCEM algorithm halts iteration of the E- and M-steps when the upper limit of a one-sided $1-\gamma$ confidence interval (CI) for the change in the Q-function rules out a preset tolerance. 

Once the MCEM algorithm has converged, uncertainty about the MLE, $\hat{\btheta}$, can be quantified using bootstrap confidence intervals or asymptotic model-based intervals, which can be obtained by inverting the observed Fisher information matrix. In the latter case, expectations of the score and Hessian required to compute the observed Fisher information can be estimated using recycling sample paths generated in the MCEM algorithm (see Section \ref{sec:mcemsupp} of the Supplementary Material). \citet{mandel2013simulation} details a convenient asymptotic Monte Carlo procedure, similar to a Bayesian posterior predictive procedure, which quantifies uncertainty for functionals of interest, such as expected sojourn times, by sampling $\bthetahat^\star$ from the asymptotic distribution of $\bthetahat$ and then simulating sample paths using $\bthetahat^\star$, which are then summarized. Monte Carlo confidence intervals are obtained by calculating quantiles of the summaries of interest. 

\subsubsection{Data-conditioned Markov proposals for semi-Markov sample paths}
\label{subsec:proposal_framework}

The challenge is to construct an efficient proposal distribution for the sample paths in the importance sampling estimate of the Q-function, \eqref{eqn:qfun}. Na\"{\i}ve forward simulation from $f_i(Z_i\mid \bX_i, \btheta)$ will produce many samples that are discordant with the data, especially as model complexity and the number of observations per participant increase. As in \citet{barone2022bayesian}, our strategy is to sample paths from a surrogate process that is easily conditioned on the data. Multistate Markov models assume that transition intensities are independent of sojourn times and are commonly used to analyze processes under intermittent observation \citep{cook2018multistate}. For our purposes, these models are attractive as surrogates, despite their unrealistic assumptions, because they are amenable to efficient algorithms for conditional simulation \citep{hobolth2009simulation}.

Let $\{\Ztil(\tau),\tau\geq 0\}$ be a continuous-time Markov process taking values in $\mcS$, and let $\bthetatil = \left(\bgammatil, \bbetatil\right)$ be a vector of Markov model parameters that we again partition into those pertaining to the baseline transition intensities and covariates. A proportional transition intensity parameterization with a time-homogeneous baseline intensity is convenient. In contrast to (\ref{eqn:semimarkovintensity}), the transition intensity is now independent of the sojourn time, $\tau_n$. Hence,
\begin{align}
    \label{eqn:markovintensity}
    \lambda_{k\ell}\left(\tau, \mcH(\tau^-), \bX(\tau^-),\bthetatil\right) = \tilde{\gamma}_{k\ell}\exp\left(\bX(\tau^-)\boldsymbol{\bbetatil}_{k\ell}\right).
\end{align} 

Suppressing the dependence on $\mcH(\tau^-)$, $\bX$ and $\bthetatil$, let $ \bLambda(\tau) = (\lambda_{k\ell}(\tau))_{k,\ell\in\mcS,} $, denote the transition intensity matrix under a Markov model, with $ \lambda_{kk}(\tau) = -\sum_{\ell \neq k}\lambda_{k\ell}(\tau) $. The transition probability matrix, $ \bP(\tau_0,\tau_1) = (p_{k\ell}(\tau_0,\tau_1))_{k,\ell\in\mcS} $, where $ p_{k\ell}(\tau_0,\tau_1) = \Pr(Z(\tau_1) = \ell \mid Z(\tau_0) = k) $, solves the Kolmogorov forward equation (KFE),
\begin{align}
	\label{eqn:kfe}
	\frac{\rmd \bP(\tau_0,\tau)}{\rmd\tau} = \bP(\tau_0,\tau)\bLambda(\tau),
\end{align}
with initial condition $ \bP(\tau_0,\tau_0) = \bI$ \citep{wilkinson2018stochastic}.  Transition probabilities can be computed via the matrix exponential, $ \bP(\tau_0,\tau_1) = \exp(\bLambda(\tau_1 - \tau_0)) $, when transition intensities are constant in $ [\tau_0,\tau_1] $, or by numerical integration of the KFE when intensities are smoothly time-varying \citep{titman2011flexible}. The KFE does not apply when intensities are semi-Markov because transition probabilities are no longer deterministic functions of parameters and initial conditions \citep{cook2018multistate}.

An individual's marginal density under a Markov model is
{\small
\begin{align}
    \label{eqn:markovlikelihood}
    r_i(\bY_i)
    &= \Pr\left(J_{i,0}\in\mcS(t_{i,0})\right)\prod_{j=1}^{L_i} \left[\Pr\left(J(t_{i,j})\in\mcS(t_{i,j})\right)^{(1 - O_{i,j})} \times \Pr\left(J(t_{i,j-1},t_{i,j})\in\mcS(t_{i,j-1},t_{i,j})\right)^{O_{i,j}}\right),\nonumber\\ 
    &= \sum_{s_0\in\mcS(t_{i,0})} \dots\sum_{s_{L_i}\in\mcS(t_{i,L_i})}\Pr(J_{i,0} = s_0)\prod_{j = 1}^{L_i}\left[p_{s_{j-1},s_j}(t_{i,j-1},t_{i,j})^{(1 - O_{i,j})}\times\right.\nonumber\\
    &\hspace{11em}\left.\left(\lambda_{s_{j-1},s_j}(t_{i,j})^{\delta_{n(t_{i,j})}}\exp\left(-\int_0^{t_{i,j} - t_{i,j-1}}\lambda_{s_{i,j-1}}(u)\rmd u\right)\right)^{O_{i,j}}\right].
\end{align}}
The likelihood for a Markov model can be maximized using standard optimization routines to yield a maximum likelihood estimate (MLE) of $\bthetatil$, under which we propose sample paths.

Algorithm \ref{alg:proposals}, diagrammed in Figure \ref{fig:samplingalgorithm}, describes how to efficiently propose sample paths conditional on the data. When the state at an observation time is not known exactly, we use the stochastic forward-filtering backwards-sampling (FFBS) algorithm to sample the sequence of state labels \citep{scott2002bayesian}. We then fill in the path over each inter-observation interval. If $Z$ is observed continuously in an interval there is no sampling to be done, and we append the observed path to our proposal. Otherwise, we sample the path in an interval using a uniformization algorithm for endpoint conditioned CTMCs \citep{hobolth2009simulation}. In the case where we know that certain states were not visited within an interval, we zero out the intensities for transitions into these states so that trajectories that visit impossible states are not proposed.

\begin{algorithmtable}
    \centering
    \renewcommand{\arraystretch}{1.5}
    \begin{tabular}{ll}
    \toprule
        \textbf{Input:} & Data: $\bY_i = \left[Y_i(t_{i,j-1},t_{i,j}),O_{i,j}\right]_{j=0,\dots,L_i}$ \\
        & Transition intensity matrices: $\left[\bLambda(t_0),\dots,\bLambda(t_{L_i-1})\right]$\\
        &  Transition probability matrices: $\left[\bP(t_{i,0},t_{i,1}), \dots,\bP(t_{i,L_i-1},t_{i,L_i})\right]$\\
        \textbf{FFBS:} & \textit{IF} any $\mcS(t_{i,j})$ or $\mcS(t_{i,j-1},t_{i,j})$ contains more than 1 state,\\
        & \hspace{1.5em} \textit{THEN} sample $\left(J_i(t_0),\dots,J_i(t_{L_i})\right)$ via FFBS.\\
        \textbf{Initialize:} & \textit{SET} $\Ztil = (J_i(t_0),0)$. \\
        \textbf{Complete path:} & \textit{FOR} $i = 1,\dots,L_i$:\\
        & \hspace{1.5em} \textit{IF} $Z_i$ is observed continuously over $[t_{i,j-1},t_{i,j}]$,\\
        & \hspace{3em} \textit{THEN} append observed path: $\Ztil \Leftarrow (\Ztil, Y_i(t_{i,j-1},t_{i,j}))$.\\
        & \hspace{1.5em} \textit{ELSEIF} $Z_i$ is observed only at $t_{i,{j-1}}$ and $t_{i,j}$,\\
        &\hspace{3em} \textit{THEN} sample $\Ztil(t_{i,j-1},t_{i,j})\mid J(t_{i,j-1}),J(t_{i,j})$ via uniformization  \\
        &\hspace{3em} \textit{AND} append sampled path: $\Ztil\Leftarrow(\Ztil, \Ztil(t_{i,j-1},t_{i,j}))$.\\
        \textbf{Return path:} & $\Ztil_i = (J, W)_i$, over $[t_{i,0}, t_{i,L_i}]$.\\
         \bottomrule
    \end{tabular}
    \caption{Procedure for drawing data-conditioned sample paths from a Markov surrogate process. The forward-filtering backward-sampling (FFBS) and uniformization algorithms are detailed in \citet{scott2002bayesian} and \cite{hobolth2009simulation}, respectively.}
    \label{alg:proposals}
\end{algorithmtable}

\subsection{Model comparison and marginal likelihood}
\label{subsec:marginalliklihood}

An advantage of the MCEM algorithm over the standard EM algorithm is the possibility to use the surrogate Markov process to construct a Monte Carlo estimator of the marginal likelihood using the data-conditioned Markov model as a proposal. The estimate of $g\left(\bY\mid \bX, \btheta^{(t)}\right)$ is
\begin{equation}  \label{eq:estimate-marginal-lik}
    \hat{g}\left(\bY\mid \bX, \btheta^{(t)}\right) = \prod_i^n r_i(\bY_i \mid \bX_i, \bthetaprime) \frac{1}{M_i}\sum_{m=1}^{M_i} \nu^{(t)}_{i,m},
\end{equation}
see Appendix~\ref{subsec:marginal-lik} for the derivation of~\eqref{eq:estimate-marginal-lik}. One can use~\eqref{eq:estimate-marginal-lik} to estimate the log-likelihood at the MLE and to estimate likelihood-based information critera, such as AIC and BIC, for model selection. Appendix~\ref{subsec:marginal-lik} details estimates for the standard error of, say, AIC, which is necessary for assessing whether an observed difference in AIC distinguishable from Monte Carlo error.

\section{Simulations}
\label{sec:simulations}

\subsection{Analysis of intercurrent events in an illness-death model}
\label{subsec:illnessdeath}
The progressive illness-death model is useful in the study of diseases, such as cancer, where death competes with disease recurrence. Patients begin follow-up when they are deemed healthy following successful treatment. The model has three states -- disease free (healthy), disease recurrence (ill), and dead -- and three transitions –- healthy to ill, healthy to dead, and ill to dead (Figure \ref{fig:illnessdeath_diagram}). Disease recurrence is interval-censored, but the time of death and its cause are observed.

We simulated clinical histories for 250 participants in 1,000 hypothetical studies using an illness-death model with Weibull transition intensities with shape parameter greater than 1, so each intensity was 0 at the time of state entry and increased over time (Tables \ref{tab:illnessdeath_pars} and \ref{tab:illnessdeath_truth}). Clinical data was accrued at baseline, which was the time of entry into the healthy state, and at random times spaced roughly every 1 or 3 months over 1 year of follow-up. The data was analyzed using a Markov model with exponential transition intensities and three semi-Markov models, which included a model with Weibull intensities and two models with B-spline intensities (see Table \ref{tab:illnnessdeath_parameterization} for parameterizations). Confidence intervals for all quantities of interest were computed by summarizing sample paths simulated from the asymptotic distribution of the MLEs. Additional details of the simulation setup and models fit are given in Appendix \ref{sec:illnessdeathsupp}. 

Table~\ref{tab:sim1res} reports bias, coverage 95\% CIs, and width of 95\% CIs for the restricted mean recurrence free survival time (RM RFST), time to recurrence (TtR) or end of follow-up (EoF), TtR among individuals whose disease recurred, and restricted mean time to death (RM TtD) following recurrence. Crude estimation by na\"{\i}vely taking the time of confirmed recurrence as the recurrence time yielded biased estimates with unacceptably low coverage, especially with less frequent follow-up. Markov models had slightly biased estimates of the RM RFST, TtR or EoF, and RM TtD. Estimates of TtR conditional on recurrence are biased and CI widths are roughly equal to the magnitude of the bias, resulting in nearly zero coverage. The bias and under-coverage are exacerbated by conditioning on recurrence because the estimate of time to recurrence preferentially involves subjects with early recurrence, which we know is unlikely under the true model.  

The three semi-Markov models all yielded unbiased estimates of the four quantities of interest. The models with Weibull intensities and linear splines with one interior knot at the median time for each transition achieved their nominal coverage, with the latter having slightly wider CIs for illness duration but similar widths for the other quantities. Natural cubic splines with two interior knots at the 1/3 and 2/3 quantiles for each transition are arguably over-parameterized for this setting, though still computationally tractable. Here, CIs were wider than those from more parsimonious models, though the discrepancy narrowed when clinical status was observed monthly.

\setcounter{table}{0}
\begin{table}
    \footnotesize\centering
        \setlength{\tabcolsep}{4pt} 
        \begin{tabular}{rrrrrrrrrrrrr}
          \toprule
          &\multicolumn{12}{c}{Follow-up every 3 months for 1 year}\\
          \cmidrule(lr){2-13}& \multicolumn{3}{c}{RM RFST} & \multicolumn{3}{c}{TtR or EoF} & \multicolumn{3}{c}{TtR $\mid$ recur.} & \multicolumn{3}{c}{RM TtD $\mid$ recur.} \\
          \cmidrule(lr){2-4} \cmidrule(lr){5-7}\cmidrule(lr){8-10}\cmidrule(lr){11-13}
          \textbf{Method} & \textbf{Bias} & \textbf{Covg.} & \textbf{CIW} & \textbf{Bias} & \textbf{Covg.} & \textbf{CIW} & \textbf{Bias} & \textbf{Covg.} & \textbf{CIW} & \textbf{Bias} & \textbf{Covg.} & \textbf{CIW} \\
  \cmidrule(lr){1-1}\cmidrule(lr){2-4} \cmidrule(lr){5-7}\cmidrule(lr){8-10}\cmidrule(lr){11-13}
  Crude est. & 0.15 & 0.05 & 0.17 & 0.1 & 0.12 & 0.12 & 0.19 & 0.03 & 0.19 & -0.25 & 0.01 & 0.21 \\
  Exponential & -0.04 & 0.89 & 0.18 & -0.02 & 0.93 & 0.14 & -0.11 & 0.0 & 0.11 & 0.01 & 0.96 & 0.22 \\
  Weibull & 0.0& 0.96 & 0.18 & 0.0& 0.95 & 0.14 & 0.0& 0.95 & 0.23 & 0.0 & 0.95 & 0.22 \\
  Linear splines & 0.0& 0.93 & 0.19 & 0.0& 0.96 & 0.15 & -0.01 & 0.95 & 0.25 & 0.0 & 0.96 & 0.31 \\
  Nat. cubic splines & 0.0& 0.86 & 0.4 & 0.0& 0.95 & 0.37 & 0.0& 0.93 & 0.44 & 0.0 & 0.97 & 0.54 \\ 
  \midrule\addlinespace[0.5em]
   &\multicolumn{12}{c}{Follow-up every month for 1 year}\\
   \cmidrule(lr){2-13}& \multicolumn{3}{c}{RM RFST} & \multicolumn{3}{c}{TtR or EoF} & \multicolumn{3}{c}{TtR $\mid$ recur.} & \multicolumn{3}{c}{RM TtD $\mid$ recur.} \\
          \cmidrule(lr){2-4} \cmidrule(lr){5-7}\cmidrule(lr){8-10}\cmidrule(lr){11-13}
          \textbf{Method} & \textbf{Bias} & \textbf{Covg.} & \textbf{CIW} & \textbf{Bias} & \textbf{Covg.} & \textbf{CIW} & \textbf{Bias} & \textbf{Covg.} & \textbf{CIW} & \textbf{Bias} & \textbf{Covg.} & \textbf{CIW} \\
            \cmidrule(lr){1-1}\cmidrule(lr){2-4} \cmidrule(lr){5-7}\cmidrule(lr){8-10}\cmidrule(lr){11-13}
  Crude est. & 0.06 & 0.75 & 0.17 & 0.04 & 0.79 & 0.13 & 0.08 & 0.67 & 0.21 & -0.1 & 0.56 & 0.22 \\
  Exponential & -0.04 & 0.91 & 0.18 & -0.02 & 0.93 & 0.14 & -0.11 & 0.01 & 0.11 & 0.01 & 0.97 & 0.22 \\
  Weibull & 0.0& 0.96 & 0.17 & 0.0& 0.95 & 0.14 & 0.0& 0.95 & 0.21 & 0.0& 0.95 & 0.2 \\
  Linear splines & 0.0& 0.95 & 0.18 & 0.0& 0.96 & 0.15 & 0.0& 0.95 & 0.23 & 0.0 & 0.96 & 0.24 \\
  Nat. cubic splines & 0.0& 0.9 & 0.24 & 0.0& 0.98 & 0.19 & 0.0& 0.96 & 0.28 & 0.0& 0.95 & 0.33 \\\bottomrule
    \end{tabular}
    \caption{Relative bias, coverage of 95\% confidence intervals, and relative confidence interval width (CIW) for estimates of restricted mean (RM) recurrence-free survival time (RM RFST), time to recurrence (TtR) or end of follow-up (EoF), TtR among individuals whose disease recurs, and RM time to death (TtD) following recurrence. Relative bias and CIW are scaled by the true value. Method refers to the baseline transition intensities or crude tabulation of events in the data.}
    \label{tab:sim1res}
\end{table}

\subsection{Estimating protective efficacy and natural history in simulated trials}
\label{subsec:sim2regen}

The analysis of the REGEN-2069 trial data in Section \ref{sec:regen2069} aims to characterize the effect of mAb on infection and seroconversion, and to quantify the duration of PCR positivity. Inference is based on the model, depicted in Figure \ref{subfig:regendiagram}, which has a na\"{\i}ve uninfected state, and four post-infection states characterized by symptom history and whether SARS-CoV-2 infection is, or has ceased to be, detectable by PCR. A two-stage procedure for calculating the probability of seroconversion given infection is detailed in Section \ref{subsubsec:regen2069asympsero}. The data consist of symptom onset times, seroconversion at four weeks, and PCR measurements roughly every week throughout the efficacy assessment period (EAP), with some variation in the visit times to emulate a realistic trial. Trials were simulated from a more complicated semi-Markov model with nine states -- a na\"{\i}ve uninfected state, and eight post-infection states characterized by symptom history, PCR positivity, and seroconversion (Table \ref{tab:regenstates}). The transitions from na\"{\i}ve to PCR+ without symptoms and from PCR+ without symptoms to diseased were assigned Weibull transition intensities, while the remainder of the transitions had exponential intensities. The shape parameter for the infection transition, $1\rightarrow 2$, was less than 1 to reflect a decreasing force of infection, and the shape parameter for symptom onset was greater than 1 to induce a delay in symptom onset. We set the simulation parameters so that subjects receiving mAb have lower rates of infection and seroconversion along with shorter durations of PCR positivity. The parameterization of the simulation model and ground truths for key quantities of interest are summarized in Tables \ref{tab:regensim_pars} and \ref{tab:regensim_truth}. 

We assessed our ability to recover key functionals with two formulations of the five state inferential model: the first had Weibull baseline intensities for transitions from na\"{\i}ve to PCR+ without symptoms ($1\rightarrow2$) and from PCR+ without symptoms to diseased ($2\rightarrow4$) and with exponential intensities from PCR+ to PCR$-$ states ($2\rightarrow3$ and $4\rightarrow5$), or another with linear spline baseline intensities with an interior knot at five days for the $1\rightarrow2$ transition and an interior knot at one week for all other transitions (Table \ref{tab:regensim_paraterization}). As benchmarks, we report crude estimates obtained by na\"{\i}vely tabulating events in the data as would be typically done in a descriptive analysis (Appendix \ref{subsubsec:crudeests}), and we also fit the 9 state simulation model to a more granular version of data where the state at each PCR assessment was known. This model is not fully identifiable with real-world data where serology is only assessed at day 28 since the order of seroconversion and cessation of detectability by PCR is unknown. Confidence intervals for all quantities were obtained via the Bayesian bootstrap \citep{rubin1981bayesian} and the bootstrap weights were stratified by treatment. Simulation results are reported in Tables \ref{tab:regensim_results}, \ref{tab:regensim_fullmod_res}, \ref{tab:regensim_parametric_res}, \ref{tab:regensim_semiparametric_res}, and \ref{tab:regensim_crude_res}.

\begin{sidewaystable}
    \small
    \begin{subtable}{\textwidth}
    \centering
    \begin{tabular}{rcccccccccccc}
        \toprule
        & \multicolumn{3}{c}{Simulation model} & \multicolumn{3}{c}{Reduced, fully par.} & \multicolumn{3}{c}{Reduced, semi par.} & \multicolumn{3}{c}{Crude estimate} \\
        \cmidrule(lr){2-4}\cmidrule(lr){5-7}\cmidrule(lr){8-10}\cmidrule(lr){11-13} & Bias & Covg. & CIW & Bias & Covg. & CIW & Bias & Covg. & CIW & Bias & Covg. & CIW\\
        \cmidrule(lr){2-4}\cmidrule(lr){5-7}\cmidrule(lr){8-10}\cmidrule(lr){11-13} PE for infec. & 0.0 & 0.96 & 0.19 & 0.0 & 0.96 & 0.19 & 0.0 & 0.96 & 0.19 & 0.13 & 0.34 & 0.2 \\
        PE for sympt. infec. & 0.0& 0.96 & 0.16 & 0.0 & 0.95 & 0.16 & 0.01 & 0.94 & 0.16 & 0.0 & 0.96 & 0.16 \\
        PE for asympt. infec. & 0.02 & 0.95 & 0.96 & -0.06 & 0.94 & 0.98 & -0.06 & 0.94 & 0.97 & 0.39 & 0.73 & 1.07 \\
        \bottomrule
    \end{tabular}
    \caption{Relative bias, coverage of 95\% confidence intervals, and relative confidence interval width for protective efficacy for infection, symptomatic infection, and asymptomatic infection. Protective efficacy is the reduction in relative risk, calculated, $PE = 1 - RR$.}
    \label{tab:regensim_pe}
    \end{subtable}
    \newline
    \vspace{1em}
    \newline
    \begin{subtable}[htbp]{\textwidth}
    \centering
    \begin{tabular}{rcccccccccccc}
        \toprule
        & \multicolumn{3}{c}{Simulation model} & \multicolumn{3}{c}{Reduced, fully par.} & \multicolumn{3}{c}{Reduced, semi par.} & \multicolumn{3}{c}{Crude estimate} \\
        \cmidrule(lr){2-4}\cmidrule(lr){5-7}\cmidrule(lr){8-10}\cmidrule(lr){11-13} & Bias & Covg. & CIW & Bias & Covg. & CIW & Bias & Covg. & CIW & Bias & Covg. & CIW\\
        \cmidrule(lr){2-4}\cmidrule(lr){5-7}\cmidrule(lr){8-10}\cmidrule(lr){11-13}RR for sympt. | infec. & 0.01 & 0.96 & 0.35 & -0.01 & 0.96 & 0.34 & -0.01 & 0.95 & 0.34 & 0.14 & 0.7 & 0.38 \\
        RR for N+ | infec. & 0.0 & 0.96 & 0.28 & 0.0 & 0.95 & 0.28 & 0.0 & 0.96 & 0.28 & 0.13 & 0.59 & 0.31 \\
        RR for N+ | sympt. & -0.01 & 0.96 & 0.3 & 0.01 & 0.96 & 0.35 & 0.02 & 0.96 & 0.36 & 0.0 & 0.96 & 0.35 \\
        RR for N+ | asympt. & 0.01 & 0.96 & 0.44 & -0.01 & 0.96 & 0.43 & -0.01 & 0.96 & 0.43 & 0.13 & 0.81 & 0.46 \\
        \bottomrule
    \end{tabular}
    \caption{Relative bias, coverage of 95\% confidence intervals, and relative confidence interval width for relative risk (RR) of symptoms following infection, and seroconversion following infection, symptomatic infection, and asymptomatic infection.}
    \label{tab:regensim_rr}
    \end{subtable}
    \newline
    \vspace{1em}
    \newline
    \begin{subtable}[htbp]{\textwidth}
    \centering
    \begin{tabular}{rcccccccccccc}
        \toprule
        & \multicolumn{3}{c}{Simulation model} & \multicolumn{3}{c}{Reduced, fully par.} & \multicolumn{3}{c}{Reduced, semi par.} & \multicolumn{3}{c}{Crude estimate} \\
        \cmidrule(lr){2-4}\cmidrule(lr){5-7}\cmidrule(lr){8-10}\cmidrule(lr){11-13} & Bias & Covg. & CIW & Bias & Covg. & CIW & Bias & Covg. & CIW & Bias & Covg. & CIW\\
        \cmidrule(lr){2-4}\cmidrule(lr){5-7}\cmidrule(lr){8-10}\cmidrule(lr){11-13}RM TI | Plac. & 0.0 & 0.95 & 0.12 & -0.02 & 0.87 & 0.12 & -0.03 & 0.85 & 0.12 & --- & --- & --- \\
        RM TI | mAb & 0.0 & 0.95 & 0.07 & -0.01 & 0.92 & 0.07 & -0.01 & 0.93 & 0.07 & --- & --- & ---\\
        RM PCR+ | Plac. & 0.0 & 0.95 & 0.11 & 0.01 & 0.94 & 0.11 & 0.02 & 0.9 & 0.11 & --- & --- & ---\\
        RM PCR+ | mAb & -0.01 & 0.96 & 0.18 & 0.01 & 0.94 & 0.18 & 0.01 & 0.94 & 0.18 & --- & --- & ---\\
        Pr(Detected | Plac.) & 0.0 & 0.96 & 0.07 & 0.0& 0.95 & 0.06 & 0.0& 0.94 & 0.07 & --- & --- & ---\\
        Pr(Detected | mAb) & 0.0 & 0.96 & 0.11 & 0.0& 0.96 & 0.1 & 0.0& 0.95 & 0.12 & --- & --- & ---\\
        \bottomrule
    \end{tabular}
    \caption{Relative bias, coverage of 95\% confidence intervals, and relative confidence interval width for restricted mean time to infection (RM TI), restricted mean duration of PCR positivity (RM PCR+), and probability of detection by PCR.}
    \label{tab:regensim_detec}
    \end{subtable}
    \caption{Evaluation of estimates of key quantities involving infection in 1,000 simulated trials. Relative bias and confidence interval width are scaled by the true value. The simulation model is described in Supplementary Appendix \ref{subsec:regen_fullmodel}. The parameterization of transition intensities in the 5 state models is given in Table \ref{tab:regensim_paraterization}. The collapsed parametric model has Weibull baseline intensities for infection and symptom onset, and exponential intensities for exiting states of detectability by PCR. The collapsed semi-parametric model has degree 1 B-splines for all baseline intensities. The simulation model was fit to data where serology was assessed weekly along with PCR, whereas the collapsed models and crude estimates used data that followed the design of REGEN-2069 where serology was assessed at day 28.}
    \label{tab:regensim_results}
\end{sidewaystable}

Crude estimates of quantities implicated by asymptomatic infection were biased and had unacceptably low coverage. The 9 state simulation model, unsurprisingly, yielded unbiased estimates and 95\% CIs that attained their nominal coverage. Estimates obtained with both of the 5-state models had negligible bias and good coverage for PE against infection, including asymptomatic infection. There was about 6\% relative bias in the PE for asymptomatic infection, which is attributable to slight underestimation of extent of asymptomatic infection on the placebo arm (Tables \ref{tab:regensim_parametric_res} and \ref{tab:regensim_semiparametric_res}). The relative risk (RR) of symptoms following infection and RR of seroconversion following infection were well estimated. Both models exhibited slight bias for the restricted mean time to infection and duration of PCR positivity along with modest under-coverage for these quantities and the probability of being detected by PCR. This is attributable to some individuals who are seropositive at the end of the EAP, but who never develop symptoms or test PCR+. The reduced model implicitly treats the timing of their infection as random from the distribution of infection times. This results in a slight bias that favors early infection under the fitted model, which is exacerbated in the placebo arm by virtue of its higher incidence of infections that are only detected by serology. Overall, the two stage estimation procedure using the reduced 5 state model has good operating characteristics for estimating key quantities of interest.

\subsection{Comparison with existing approaches}
\label{subsec:comparison_main}
In Section \ref{subsec:araliscomp} of the Supplementary Material, we compare our approach with that taken in \citet{aralis2019stochastic}, who generate paths from a semi-Markov model by forward simulation and reject proposed paths that are inconsistent with the data. We explore the performance of the two approaches for fitting a progressive illness-death model and find that data-conditioned path proposals yield dramatic efficiency gains as the number of observations per-individual increases (Table \ref{tab:comparison-aralis}). The relative efficiency gains of our approach are greater for more complex models since the rejection sampler becomes increasingly likely to propose paths that are inconsistent with the data.

In Section \ref{subsec:phasetype_comp} of the Supplementary Material, we compare semi-parametric models fit using our approach with phase-type models that are commonly used to approximate semi-Markov models in panel data settings. This latter class of models uses a latent Coxian phase-type structure to approximate arbitrary time-to-event distributions and is amenable to numerical optimization \citep{titman2010semi}. Phase-type models can struggle to approximate intensities that are initially close to zero and remedies for ameliorating the ensuing lack of fit introduce additional technical complications and computational challenges. We demonstrate that approximating the transition intensity for illness-onset in a semi-Markov illness-death process using a B-spline with a single interior knot yields a substantially improved fit compared with a phase-type model with two latent states to capture the semi-Markov dynamics of this transition (Table \ref{tab:comparison-phase-type}). In particular, expected prevalence of illness is too early and too low under the phase-type model, whereas the semi-Markov model accurately tracks the true illness prevalence over time (Figure \ref{fig:phase-type-prevalence}). The phase-type model still vastly outperforms a simple Markov model and we speculate that phase-type models could be used as an alternative proposal distribution within our MCEM framework in settings where simple time-homogeneous Markov models are inefficient. Phase-type models are characterized by a latent Markovian structure that, like simpler Markov dynamics, lends itself to data-conditioned simulation. Whether a more complex alternative proposal distribution is warranted could be determined based on having a small ESS relative to the nominal number of paths.  

\section{Protective efficacy and natural history in REGEN-2069}
\label{sec:regen2069}

\subsection{Trial of REGEN-COV mAb for COVID-19 prophylaxis} 
\label{subsec:regen2069descrip2}
The primary analysis of Part A of the REGEN-2069 trial studied occurrence of symptomatic infection. \citep{obrien2021regen}. Participants randomized to REGEN-COV had 81.4\% lower risk of symptomatic infection within the EAP compared with participants in the placebo arm (odds ratio, 0.17; 95\% CI, 0.09 to 0.33; p $<$ 0.001). We have several objectives in our reanalysis of REGEN-2069: First, we aim to estimate incidence of infection over the EAP and to quantify the PE of mAb prophylaxis against infection, including asymptomatic infections. Understanding whether the mAb effect on asymptomatic infection is similar to that on symptomatic infection is critical to short-circuiting transmission within households. Second, we estimate the effect of mAb on seroconversion and whether this effect varies by symptomatic and asymptomatic infection. Finally, we attempt to unravel how the effect of mAb might have itself affected detectection of asymptomatic infections that are incompletely captured with episodic RT-qPCR sampling and serological testing. To this end, we estimate the duration of detectable viral shedding and estimate the probability that an infection would have been detected on the basis of PCR alone at a scheduled assessment. 

Figure~\ref{fig:regen_consort} summarizes the data processing criteria used to create our analysis dataset, which differs slightly from the dataset used in the primary analysis in \citet{obrien2021regen}, and Table~\ref{tab:regen_summary} tabulates outcomes by arm. Our dataset recorded 9 cases of COVID-19 among the 792 participants in the mAb arm and 56 cases out of 796 participants in the placebo arm. In the mAb arm, there were an additional 37 asymptomatic infections, 9 of which were detected by serology and 28 by RT-qPCR. In the placebo arm, there were 61 asymptomatic infections, 27 of which were positive by RT-qPCR and serology, 22 of which were positive by RT-qPCR alone, and 12 of which were positive by serology. In crude terms, asymptomatic infection was more common, and seroconversion was less common, among participants treated with mAb. In their analysis of REGEN-2069, \citet{follmann2022antinucleocapsid} suggest that a likely explanation for this lower rate of seroconversion is a mAb-induced reduction in viral loads, which in turn suppresses symptoms and reduces exposure to N-antigen.

\subsection{Model for clinical and immunological progression}
\label{subsec:regen2069model}
We model the progression of participants through five states -- a na\"{\i}ve uninfected state at enrollment and four post-infection characterized by symptom history and whether SARS-CoV-2 infection is, or has ceased to be, detectable by PCR. The assumptions under which the model is formulated are summarized in Table \ref{tab:regenassump} and the model transitions are depicted in Figure \ref{subfig:regendiagram}). In keeping with published findings about the kinetics of viral shedding and COVID-19 progression \citep{puhach2023sars}, we assume that PCR positivity precedes symptom onset and that participants no longer become symptomatic after they have stopped shedding detectable virus. Thus, participants who become PCR+ (state 2) either remain asymptomatic and transition to PCR$-$ (state 3), or become symptomatic (state 4) and then PCR$-$ (state 5). Seroconversion is not modeled as an explicit transition, but nonetheless provides information about whether a participant was infected. A positive serology for a participant who was consistently PCR$-$ with no symptoms implies that they were a short shedder and could have been in either of state 1 or state 3 at any interim PCR assessment. This means that such a participant's state needs to be sampled at PCR assessment times using the FFBS algorithm within the path proposal framework (Algorithm \ref{alg:proposals} and Figure \ref{fig:samplingalgorithm}).

We fit $6$ models, summarized in Tables \ref{tab:regen-aic} and \ref{tab:5state_parameterization}, that were distinguished by the parameterization of their baseline transition intensities, i.e., $\eta_{k\ell}(\tau_n,\bgamma_{k\ell})$ in equation (\ref{eqn:semimarkovintensity}). The most flexible model used piecewise linear baseline intensities with 3 interior knots at 3.5, 10.5, and 17.5 days for the transitions from na\"{\i}ve to PCR+ and from PCR+ to symptomatic, and 1 interior knot at 1 week for each of the PCR+ to PCR- transitions. A Markov model with time-homogeneous exponential intensities was the most rigid fully-parametric model. All intensities were adjusted for mAb assignment via a proportional intensity parameterization.

We directly maximized the likelihood for the Markov model and used the MCEM algorithm for the remaining semi-Markov models. We screened the candidate models using the Aikaike information criterion (AIC; \citet{akaike1998information}), which has been applied in model selection with multistate Markov models fit to interval-censored data \citep{machado2021penalised}. Estimates for the quantities of interest in Figure~\ref{fig:regen_estimates} were computed by Monte Carlo and confidence intervals were computed via the Bayesian bootstrap \citep{rubin1981bayesian}. The bootstrap weights were stratified on treatment arm to preserve the balance of treatment assignments. 

\subsubsection{Seroconversion and asymptomatic infection}
\label{subsubsec:regen2069asympsero}
Let $C_A$, $C_S$ and $C=C_A+C_S$ denote the number of asymptomatic, symptomatic, and total seroconversions observed in an arm. Also let  $\hat{I}_A$, $\hat{I}_S$, and $\hat{I}$ be the analogous number of infections estimated by the model. The estimated probabilities of seroconversion are $\hat{V}_A = C_A / \hat{I}_A$, $\hat{V}_S = C_S / \hat{I}_S$, and $\hat{V} = C / \hat{I}$.  Note that we use the estimated probability of asyptomatic infection since otherwise $\hat{I}_S + \hat{I}_A$ may not equal $\hat{I}$, which would not be coherent under the model. Our use of the bootstrap is motivated by the observation that model-based asymptotic confidence intervals will understate the uncertainty in seroconversion following asymptomatic infection by virtue of conditioning on the number of asymptomatic seroconversion events. 

\subsection{Results}
\label{subsec:regen2069results}

Table~\ref{tab:regen-aic} compares the fitted models in terms of AIC. To showcase the improvement of the semi-Markov models over the Markov model, the difference in log-likelihood with the Markov model is displayed. The semi-Markov models offer a gain in log-likelihood between 78.7 and 101 units and the trade-off at the cost of additional model parameters is worthwhile, as reflected in the smaller AIC for more flexible models. The preferred model by AIC was parameterized with degree 1 B-splines with 2 interior knots for the $1\rightarrow2$ and $2\rightarrow4$ transitions, and 1 interior knot for the $2\rightarrow 3$ and $4\rightarrow 5$ transitions. The Markov model exponential baseline intensities had, by far, the worst AIC, while the model with Weibull intensities was worst among the semi-Markov models.

Figures \ref{subfig:incidence-disease} and \ref{subfig:incidence-infection} display the pointwise estimates and 95\% confidence bands for cumulative incidence of symptomatic infection and all-comer infection, including asymptomatic infection, obtained using the best fit model. The observed incidence of symptomatic infections, which is continuously observed, is fully captured by the confidence bands, an indication that the semi-Markov model is a good fit. We observe a large increase in incidence of infections on day 7 (Figure~\ref{subfig:incidence-infection}) --- these are participants who were infected early, so are pre-/asymptomatic, and who have not yet been identified as infected. Thus, the model smoothly estimates a sharp increase in infections shortly after randomziation and avoids the artifactual bump at day 7 that is an artifact of weekly PCR assessment. The discrepancy between the observed and estimated incidence is greatest in the first week, when the majority of infections occurred --- 64.5\% (95\% CI: 57.2\%, 72.0\%) of the total incidence in the placebo arm and 63.3\% (95\% CI: 55.6\%, 71.3\%) in the mAb arm. Furthermore, the model has higher cumulative incidence at day 28, which reflects infections of short duration that are missed by intermittent sampling. 

In our dataset, REGEN-COV prophylaxis reduced the risk of symptomatic COVID-19 by 83.6\% (95\% CI: 69.4\%, 93.1\%; Figure \ref{fig:regen_estimates}), consistent with the estimate of 81.4\% reduction in \citet{obrien2021regen}. We estimate that 14.6\% (95\% CI: 12.1\%, 17.3\%) of participants in the placebo arm were infected, symptomatically or asymptomatically, by the end of the EAP as compared with 5.7\% (95\% CI: 4.2\%, 7.6\%) of participants in the mAb arm. Overall PE against all-comer infection was 60.4\% (95\% CI: 44.9\%, 72.5\%) and PE against asymptomatic infection, i.e., the reduction in the probability of infection that are not symptomatic in the EAP, was 38.7\% (95\% CI: 10.0\%, 60.8\%). Participants who received mAb were less likely to develop symptoms following infection (RR, 41.2\%; 95\% CI: 18.9\%, 67.7\%). Symptomatic infections were a larger fraction of the overall infection burden in the placebo arm than in the mAb arm, 47.5\% (95\% CI: 38.1\%, 56.9\%) versus 19.4\% (95\% CI: 9.4\%, 31.7\%), respectively. 

Infected participants on the mAb arm were 31.9\% (95\% CI: 22.3\%, 44.6\%) as likely to seroconvert following infection compared with participants on the placebo arm (Figure \ref{subfig:whiskers}). The rate of seroconversion by day 28 among asymptomatically infected placebo participants was lower than among symptomatically infected placebo participants, 64.3\% (95\% CI: 52.2\%, 77.7\%) versus 77.5\% (95\% CI: 65.5\%, 88.3\%), but significantly higher than the rate of seroconversion among asymptomatic mAb participants, 24.6\% (95\% CI: 12.2\%, 40.6\%). This dynamic lends support to the hypothesis that high viral loads increase both the severity of disease and the intensity of the ensuing immune response leading to expression of anti-nucleocapsid antibodies. 

Finally, we find that REGEN-COV reduces the mean duration of detectable viral shedding from 13.0 days (95\% CI: 11.5 days, 14.6 days) to 6.2 days (95\% CI: 5.0 days, 7.8 days). The estimated durations of PCR positivity are shorter than the estimates reported in \citet{obrien2021regen} since we allow for seropositive participants to contribute information about the duration of viral shedding, even if they never develop symptoms or test positive by PCR. One obvious consequence of mAb participants having a shorter duration of PCR positivity is that the probability of detecting infection by PCR alone, i.e., without appealing to symptoms or serology and assuming that PCR assessments were conducted exactly weekly, is lower on the mAb arm; 0.69 (95\% CI: 0.58, 0.78) versus 0.90 (95\% CI: 0.85, 0.94). Thus, data assimilation is critical to capturing the full extent of infection since mAb reduces the sensitivity of symptoms and PCR for identifying cases.

\begin{figure}[htbp]
    \centering
    \begin{subfigure}[b]{.475\textwidth}
        \includegraphics[width=\textwidth]{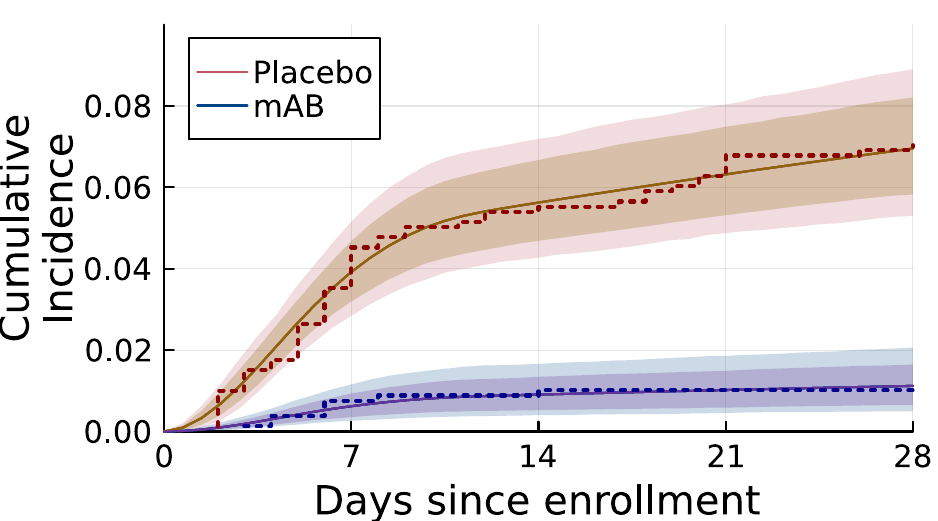}
        \captionsetup{skip=0.5em}
        \caption{Incidence of confirmed COVID-19.}
        \label{subfig:incidence-disease} 
        \vspace{2em} 
        \includegraphics[width=\textwidth]{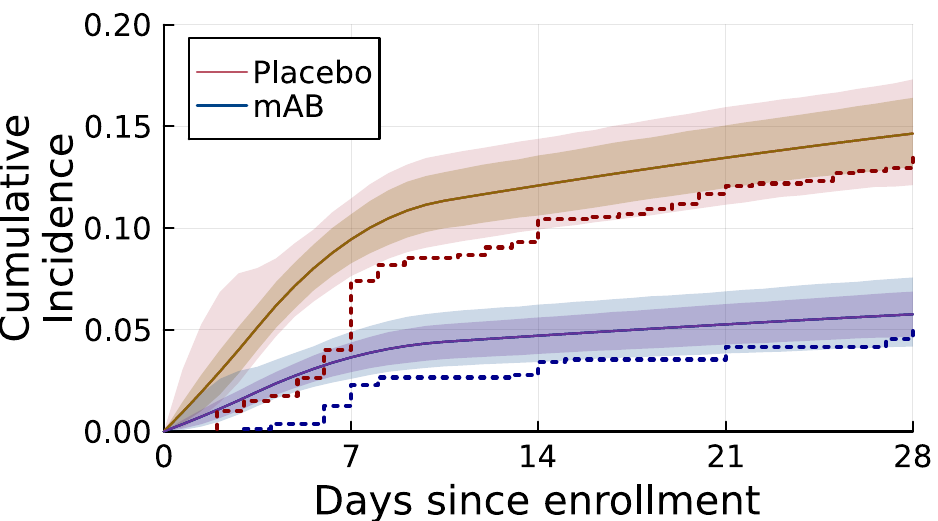}
        \captionsetup{skip=0.5em}
        \caption{Incidence of all-comer infection.}
        \label{subfig:incidence-infection} 
    \end{subfigure}
    \begin{subfigure}[b]{.475\textwidth}
        \includegraphics[width=\textwidth]{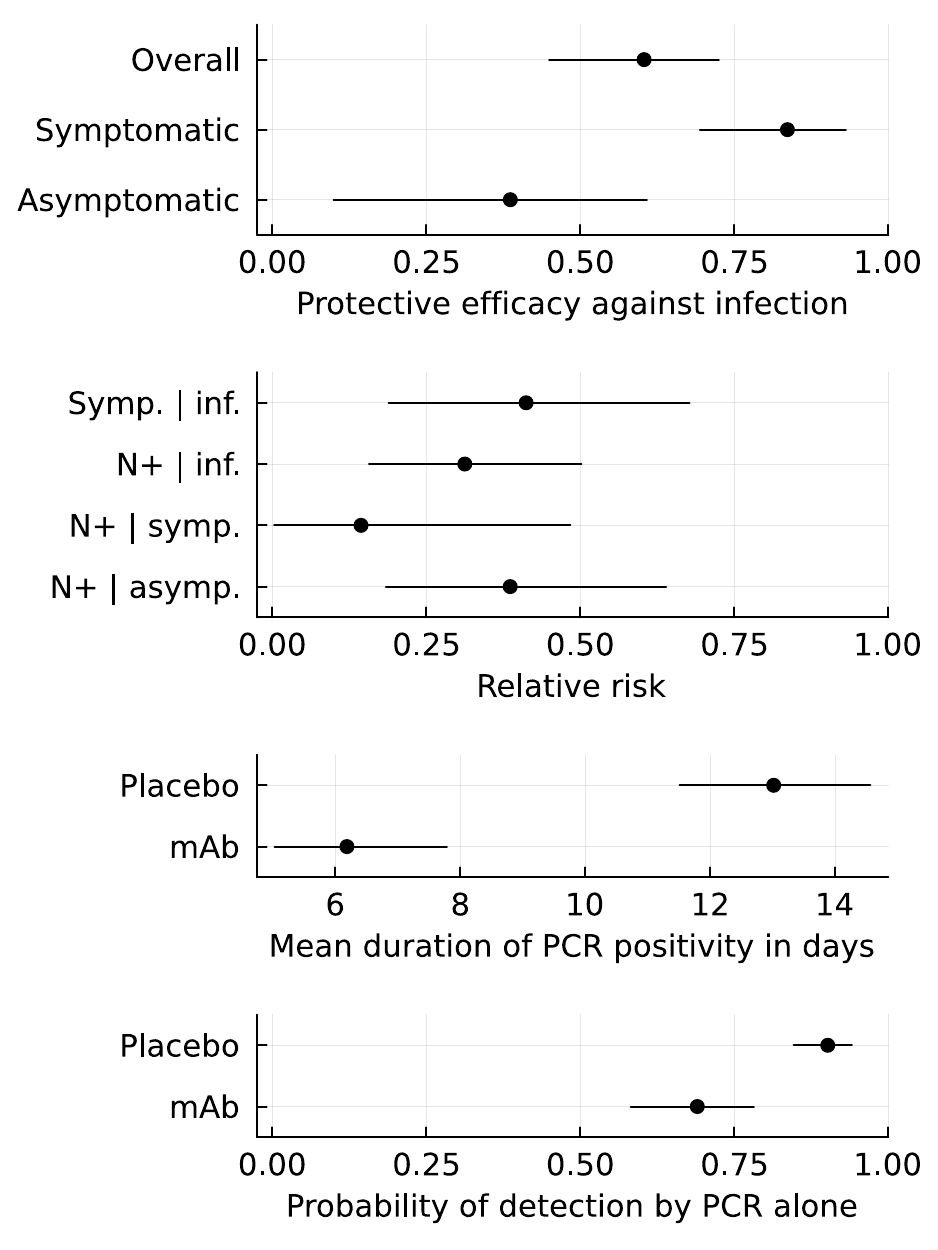}
        \captionsetup{skip=0pt} 
        \caption{Quantities of interest.}
        \label{subfig:whiskers}
    \end{subfigure}
    \caption{Cumulative incidence of disease (\ref{subfig:incidence-disease}) and infection (\ref{subfig:incidence-infection}) by arm, estimates of key functionals of interest, and confidence intervals based on 1,000 bootstrap samples. Solid lines correspond to the pointwise estimates, the shaded areas are $80\%$ and $95\%$ pointwise confidence band, and the dotted lines are the observed incidence. In Panel \ref{subfig:whiskers}, the dots are the maximum likelihood estimates and the bars are $95\%$ confidence intervals. Additional detailed results are reported in Table \ref{tab:regen_fullres}.}
    \label{fig:regen_estimates}
\end{figure}

\section{Discussion}
\label{sec:discussion}

We combined PCR, symptom, and serology data to study the effect of mAb on difficult to measure endpoints implicated by asymptomatic infection. Our findings support a multifaceted benefit of mAb prophylaxis -- it reduces the risk of infection, lessens the likelihood of developing symptoms following infection, and shortens the duration of viral shedding. We find that the data are consistent with a reduction in the risk of asymptomatic infection and that this reduction is modest compared with the reduction in the risk of symptomatic infection. We conclude that mAb prophylaxis could serve dual purposes of infection control and shielding vulnerable individuals from disease once infected. Our analysis also finds that mAb likely reduces the likelihood of seroconversion among the asymptomatically infected, supporting results in \citet{follmann2022antinucleocapsid}.

Multistate semi-Markov models encode more realistic biological assumptions than Markov models that have historically been applied in analyzing data under panel observation. Our computational framework is general, rather than model specific, and allows for misclassification in the observed state. This innovation is critical to our analysis of the REGEN-2069 data. We employ importance sampling to recycle sample paths within an MCEM algorithm, thus minimizing computational overhead. We further explore semi-parametric inference via splines and leave the door open for further work on penalized splines. It would be interesting to explore likelihoods that jointly penalize the flexibility of all semi-parametric components, as in \citet{machado2021penalised}. We also regard optimization-centric Bayesian approaches, such as the weighted Bayesian bootstrap \citep{newton2021weighted}, to inference for multistate models in biomedical applications as an important avenue for future work since there is a need to use prior information to resolve aspects of the model that are not identified in the data. Our framework would naturally fold into such an approach, which turns the problem of posterior sampling into a set of optimization problems that can be solved in parallel while allowing for prior information about observables to be encoded via penalized likelihoods or pseudo-observations.

Our computational framework is closest, in spirit, to the Bayesian data augmentation MCMC algorithm of \citet{barone2022bayesian}, who also employ a Markov surrogate to propose semi-Markov sample paths conditionally on the data. While data augmentation MCMC may be tractable for fitting simple models to small datasets, it is difficult to see this as a general solution in more complicated modeling scenarios when it becomes increasingly expensive to sample paths at each MCMC iteration. Furthermore, convergence and mixing deteriorate as the data become increasingly sparse relative to dynamics of the process \citep{papaspiliopoulos2007general}. These problems are compounded when poor convergence and mixing necessitate longer MCMC runs to obtain an adequate posterior sample. 

We caution that our MCEM algorithm is not a panacea for non-identifiability despite its being agnostic about the structure of the model and accommodating a variety of measurement processes. General conditions under which MLEs for intermittently observed Markov processes may fail to exist, or are practically not identifiable, are characterized in \citet{bladt2005statistical}. Their conclusions, such as needing sufficient temporal granularity in the data to ensure practical identifiability, apply equally to semi-Markov models. Scenarios that are computationally challenging for maximum likelihood estimation often prove difficult for Bayesian computation in the absence of well-calibrated priors. In our experience, our algorithm tends to fail ``loudly" in settings where a responsible data analyst would reasonably expect that a model is over-parameterized given the extent of the data. We view obvious computational failure as a strength of the algorithm since it indicates that the model is not supported by the data.

We have limited our exposition to multistate models that are progressive and where the observation process is independent of the underlying disease process, both of which are reasonable in the context the REGEN-2069 trial. It is possible, in theory, to apply our MCEM algorithm to models with back-transitions. In practice, fitting models for recurrent processes under intermittent observation is difficult and often requires knowledge about the number of back-transitions to be brought to bear by restricting the state space of the model or by appealing to informative priors or a penalized likelihood. Thus, we defer exploration of recurrent models for future work. We also highlight extensions towards automatic smoothing parameter selection in penalized maximum likelihood estimation as being of particular importance given the need to mitigate the influence of modeling choices that could have undue influence on model fit, e.g., knot placement for regression splines. Penalized maximum likelihood estimation was explored in the Markov case in \citet{machado2021penalised} and it would be of great interest to adapt aspects of their approach to fitting semi-Markov models via MCEM.

It is also commonly recognized that failing to accommodate for non-ignorability of the observation process, as would arise under a disease driven observation (DDO) processes, can confound inference \citep{lange2018estimating, cook2021independence}. It would be straightforward to extend our framework to accommodate DDO processes by modifying the likelihood and proposal algorithm to account for preferential sampling. The proposal algorithm could also be easily modified to accommodate more complicated observation schemes, such as when data indicate that a state was not visited over some time interval, e.g., in the ACTT-1 study of remdesivir for treatment of COVID-19 the data recorded the highest-intensity level of respiratory therapy over the preceding 24 hours. This could be addressed by setting intensities to zero for transitions into states that are ruled out by the data, thus assuring that impossible trajectories would not be considered. Finally, our MCEM algorithm can also be extended by substituting the Markov proposal model with a richer proposal model from which it is straightforward to sample paths conditional on the panel data. Candidates include discrete mixtures of Markov models \citep{elvira2019generalized} and phase-type continuous-time Markov chains \citep{titman2010semi}. These richer proposal would have the potential of making the importance sampling estimate of the marginal log-likelihood, and thereby the MCEM algorithm as a whole, more robust.

\section{Software}
\label{sec:software}
The methods presented in this manuscript are implemented in the \texttt{MultistateModels.jl} Julia package, available at \url{https://github.com/fintzij/MultistateModels.jl}. An `R` package making the Julia software callable from `R` is available at \url{https://github.com/ammateja/multistatemodels}  together with vignettes demonstrating how to simulate from and fit various multistate models. Archived versions of these packages and code that can be used to reproduce this manuscript are available at \url{https://github.com/ammateja/MultistateModelsPaper}.

\section{Supplementary Material}
\label{sec:supplement} 
Supplementary material is available online at
\url{http://biostatistics.oxfordjournals.org}. The supplementary material includes appendices expanding on technical aspects of the methods in this manuscript, details and additional results for simulations, and details about data pre-processing and model parameterization for the REGEN-2069 analysis, as well as additional results. 

\section*{Disclaimer}
The content of this publication does not necessarily reflect the views or policies of the Department of Health and Human Services and the Food and Drug Administration, nor does mention of trade names, commercial products, or organizations imply endorsement by the U.S. Government.

\section*{Acknowledgments}
The authors would like to thank to Michael P. Fay, Jing Wang, Arthur Winer, and Jennifer Gillman for their invaluable help that greatly improved this manuscript. This work utilized the computational resources of the NIH HPC Biowulf cluster (https://hpc.nih.gov).
This project has been funded in whole or in part with federal funds from the National Cancer Institute, National Institutes of Health, under Contract No. 75N91019D00024. 
{\it Conflict of Interest}: None declared.

\bibliographystyle{biorefs}
\bibliography{refs}


\newpage
\setcounter{figure}{0}
\renewcommand{\thefigure}{S\arabic{figure}}
\setcounter{equation}{0}
\renewcommand{\theequation}{S\arabic{equation}}
\setcounter{table}{0}
\renewcommand{\thetable}{S\arabic{table}}
\appendix

\newpage
\section{Technical Appendix: Additional methodological details}
\label{sec:mcemsupp}

\subsection{Graphical representation of the sampling algorithm}
\label{subsec:samplingdiagram}
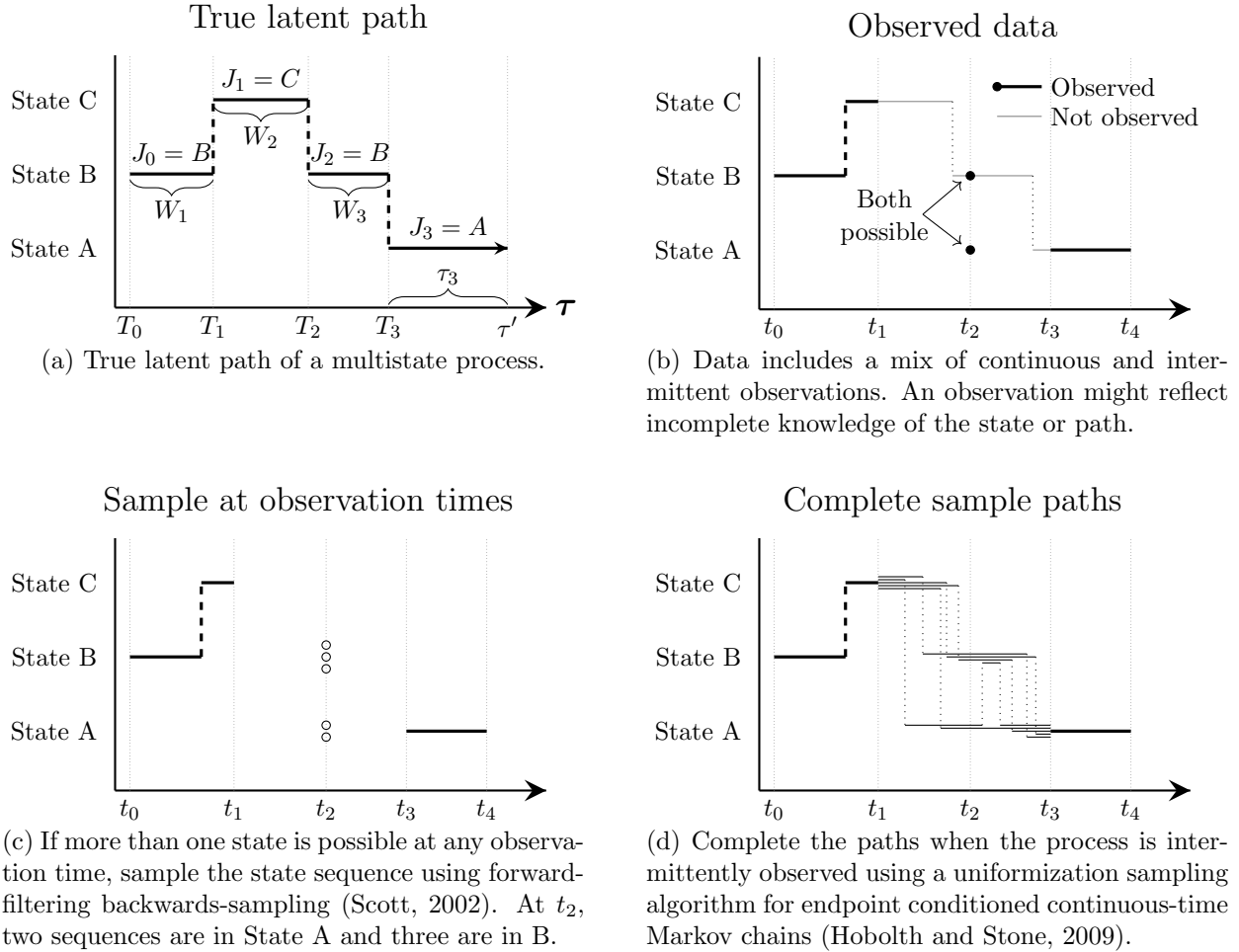
\begin{figure}[htbp]
\centering %
\setkeys{Gin}{width=\linewidth}
    \begin{subfigure}[t]{0.475\textwidth}
        \begin{tikzpicture}[scale=0.8,every node/.style={scale=0.9}]
        \definecolor{cbgray1}{gray}{0.5}
        \definecolor{cbgray2}{gray}{0.25}

        \draw (3, 3.2) node[anchor=south] {\Large True latent path};

        \draw [thick] (-0.25,-1.25) -- (-0.25,3);
        \draw[decoration={markings,mark=at position 1 with {\arrow[scale=2,>=stealth]{>}}},postaction={decorate},thick] (-0.25,-1.25) -- (7,-1.25);
        \draw (7, -1.25) node[anchor=west] {\Large$\boldsymbol{\tau}$};

        \foreach \i in {0.0, 1.4, 3, 4.35, 6.35} \draw [very thin, densely dotted, color = cbgray1] (\i,-1.25)--(\i,3);
        
        \draw (0.0, -1.25) node[anchor = north] {$T_0$};
        \draw (1.4, -1.25) node[anchor = north] {$T_1$};
        \draw (3, -1.25) node[anchor = north] {$T_2$};
        \draw (4.35, -1.25) node[anchor = north] {$T_3$};
        \draw (6.35, -1.25) node[anchor = north] {$\tau^\prime$};

        \draw (-0.4, 2.25) node[anchor = east] {State C};
        \draw (-0.4, 1) node[anchor = east] {State B};
        \draw (-0.4, -0.25) node[anchor = east] {State A};

        \draw [very thick] (0.0, 1) -- (1.4, 1);
        \draw [very thick] (1.4, 2.25) -- (3, 2.25);
        \draw [very thick] (3, 1) -- (4.35, 1);
        \draw [very thick, decoration={markings,mark=at position 1 with {\arrow[scale=1,>=stealth]{>}}},postaction={decorate}] (4.35, -0.25) -- (6.35, -0.25);

        \draw [very thick, dashed] (1.4, 1) -- (1.4, 2.25);
        \draw [very thick, dashed] (3, 2.25) -- (3, 1);
        \draw [very thick, dashed] (4.35, 1) -- (4.35, -0.25);

        \draw (0.7, 1) node[anchor = south] {$J_0 = B$};
        \draw (0.7, 0.7) node[anchor = north] {$W_1$};
        \draw [decorate,decoration={brace,amplitude=6pt,mirror}] (0.0,0.9) -- (1.4,0.9);

        \draw (2.2, 2.25) node[anchor = south] {$J_1 = C$};
        \draw (2.2, 1.95) node[anchor = north] {$W_2$};
        \draw [decorate,decoration={brace,amplitude=6pt,mirror}] (1.4,2.15) -- (3,2.15);

        \draw (3.7, 1) node[anchor = south] {$J_2 = B$};
        \draw (3.7, 0.7) node[anchor = north] {$W_3$};
        \draw [decorate,decoration={brace,amplitude=6pt,mirror}] (3,0.9) -- (4.35,0.9);

        \draw (5.35, -0.25) node[anchor = south] {$J_3 = A$};
        \draw (5.35, -0.45) node[anchor = north] {$\tau_{3}$};
        \draw [decorate,decoration={brace,amplitude=6pt}] (4.35,-1.2) -- (6.35,-1.2);

    \end{tikzpicture}
    \caption{True latent path of a multistate process.}
    \label{fig:processexample2}
    \end{subfigure}
    \hfill
    \begin{subfigure}[t]{0.475\textwidth}
        \begin{tikzpicture}[scale=0.8,every node/.style={scale=0.9}]
        \definecolor{cbgray1}{gray}{0.5}
        \definecolor{cbgray2}{gray}{0.25}

        \draw (3, 3.2) node[anchor=south] {\Large Observed data};

        \draw [thick] (-0.25,-1.25) -- (-0.25,3);
        \draw[decoration={markings,mark=at position 1 with {\arrow[scale=2,>=stealth]{>}}},postaction={decorate},thick] (-0.25,-1.25) -- (7,-1.25);

        \foreach \i in {0.0, 1.75, 3.3, 4.65, 6} \draw [very thin, densely dotted, color = cbgray1] (\i,-1.25)--(\i,3);
        \draw (0.0, -1.25) node[anchor = north] {$t_0$};
        \draw (1.75, -1.25) node[anchor = north] {$t_1$};
        \draw (3.3, -1.25) node[anchor = north] {$t_2$};
        \draw (4.65, -1.25) node[anchor = north] {$t_3$};
        \draw (6, -1.25) node[anchor = north] {$t_4$};

        \draw (-0.4, 2.25) node[anchor = east] {State C};
        \draw (-0.4, 1) node[anchor = east] {State B};
        \draw (-0.4, -0.25) node[anchor = east] {State A};

        \draw [very thick] (0.0, 1) -- (1.2, 1);
        \draw [very thick] (1.2, 2.25) -- (1.75, 2.25);
        \draw [very thin, color=cbgray1] (1.75, 2.25) -- (3, 2.25);
        \draw [very thin, color=cbgray1] (3, 1) -- (4.35, 1);
        \draw [very thin, color=cbgray1] (4.35, -0.25) -- (4.65, -0.25);
        \draw [very thick] (4.65, -0.25) -- (6, -0.25);

        \draw [very thick, dashed] (1.2, 1) -- (1.2, 2.25);
        \draw [thin, dotted] (3, 2.25) -- (3, 1);
        \draw [thin, dotted] (4.35, 1) -- (4.35, -0.25);

        \filldraw[black] (3.3,1) circle (2pt) node {};
        \filldraw[black] (3.3, -0.25) circle (2pt) node {};


        \draw [thin, decoration={markings,mark=at position 1 with {\arrow[scale=1]{>}}},postaction={decorate}] (2.5, 0.35) -- (3.15, -0.15);
        \draw [thin, decoration={markings,mark=at position 1 with {\arrow[scale=1]{>}}},postaction={decorate}] (2.5, 0.35) -- (3.15, 0.9);
        \draw (2.75, 0.3) node [anchor=east,align=center] {Both \\ possible};

        \filldraw (3.75, 2.5) circle (2pt) node {};
        \draw [very thick] (3.75, 2.5) -- (4.5, 2.5); 
        \draw (4.5, 2.5) node [anchor=west] {Observed};
        
        \draw [thin, color=cbgray1] (3.75, 2) -- (4.5, 2); 
        \draw (4.5, 2) node [anchor=west] {Not observed};
        \end{tikzpicture}
        \caption{Data includes a mix of continuous and intermittent observations. An observation might reflect incomplete knowledge of the state or path.}
        \label{fig:dataexample2}
    \end{subfigure}

    \ContinuedFloat\vskip\baselineskip
    \begin{subfigure}[t]{0.475\textwidth}
        \begin{tikzpicture}[scale=0.8,every node/.style={scale=0.9}]
        \definecolor{cbgray1}{gray}{0.5}
        \definecolor{cbgray2}{gray}{0.25}

        \draw (3, 3.2) node[anchor=south] {\Large Sample at observation times};

        \draw [thick] (-0.25,-1.25) -- (-0.25,3);
        \draw[decoration={markings,mark=at position 1 with {\arrow[scale=2,>=stealth]{>}}},postaction={decorate},thick] (-0.25,-1.25) -- (7,-1.25);

        \foreach \i in {0.0, 1.75, 3.3, 4.65, 6} \draw [very thin, densely dotted, color = cbgray1] (\i,-1.25)--(\i,3);
        \draw (0.0, -1.25) node[anchor = north] {$t_0$};
        \draw (1.75, -1.25) node[anchor = north] {$t_1$};
        \draw (3.3, -1.25) node[anchor = north] {$t_2$};
        \draw (4.65, -1.25) node[anchor = north] {$t_3$};
        \draw (6, -1.25) node[anchor = north] {$t_4$};

        \draw (-0.4, 2.25) node[anchor = east] {State C};
        \draw (-0.4, 1) node[anchor = east] {State B};
        \draw (-0.4, -0.25) node[anchor = east] {State A};

        \draw [very thick] (0.0, 1) -- (1.2, 1);
        \draw [very thick] (1.2, 2.25) -- (1.75, 2.25);
        \draw [very thick] (4.65, -0.25) -- (6, -0.25);

        \draw [very thick, dashed] (1.2, 1) -- (1.2, 2.25);

        \draw[black] (3.3,1.2) circle (2pt) node {};
        \draw[black] (3.3,1.0) circle (2pt) node {};
        \draw[black] (3.3,0.8) circle (2pt) node {};
        \draw[black] (3.3, -0.15) circle (2pt) node {};
        \draw[black] (3.3, -0.35) circle (2pt) node {};

        
        \end{tikzpicture}
        \caption{If more than one state is possible at any observation time, sample the state sequence using forward-filtering backwards-sampling \citep{scott2002bayesian}. At $t_2$, two sequences are in State A and three are in B.}
        \label{fig:possiblepathsobstimes}
    \end{subfigure}
    \hfill
    \begin{subfigure}[t]{0.475\textwidth}
        \begin{tikzpicture}[scale=0.8, every node/.style={scale=0.9}]
        \definecolor{cbgray1}{gray}{0.5}
        \definecolor{cbgray2}{gray}{0.25}

        \draw (3, 3.2) node[anchor=south] {\Large Complete sample paths};

        \draw [thick] (-0.25,-1.25) -- (-0.25,3);
        \draw[decoration={markings,mark=at position 1 with {\arrow[scale=2,>=stealth]{>}}},postaction={decorate},thick] (-0.25,-1.25) -- (7,-1.25);

        \foreach \i in {0.0, 1.75, 3.3, 4.65, 6} \draw [very thin, densely dotted, color = cbgray1] (\i,-1.25)--(\i,3);
        \draw (0.0, -1.25) node[anchor = north] {$t_0$};
        \draw (1.75, -1.25) node[anchor = north] {$t_1$};
        \draw (3.3, -1.25) node[anchor = north] {$t_2$};
        \draw (4.65, -1.25) node[anchor = north] {$t_3$};
        \draw (6, -1.25) node[anchor = north] {$t_4$};

        \draw (-0.4, 2.25) node[anchor = east] {State C};
        \draw (-0.4, 1) node[anchor = east] {State B};
        \draw (-0.4, -0.25) node[anchor = east] {State A};

        \draw [very thick] (0.0, 1) -- (1.2, 1);
        \draw [very thick] (1.2, 2.25) -- (1.75, 2.25);
        \draw [very thick] (4.65, -0.25) -- (6, -0.25);

        \draw [very thick, dashed] (1.2, 1) -- (1.2, 2.25);

        \draw [thin] (1.75, 2.35) -- (2.5, 2.35); 
        \draw [thin] (1.75, 2.3) -- (2.2, 2.3); 
        \draw [thin] (1.75, 2.25) -- (2.9, 2.25);
        \draw [thin] (1.75, 2.2) -- (3.1, 2.2); 
        \draw [thin] (1.75, 2.15) -- (2.8, 2.15);

        \draw [thin, dotted] (2.5, 2.35) -- (2.5, 1.05);
        \draw [thin, dotted] (2.9, 2.25) -- (2.9, 1.0);
        \draw [thin, dotted] (3.1, 2.2) -- (3.1, 0.95);
        \draw [thin, dotted] (2.2, 2.3) -- (2.2, -0.15);
        \draw [thin, dotted] (2.8, 2.15) -- (2.8, -0.2);

        \draw [thin] (2.5, 1.05) -- (4.25, 1.05);
        \draw [thin] (2.9, 1.0) -- (4.4, 1.0);
        \draw [thin] (3.1, 0.95) -- (4.0, 0.95);
        \draw [thin] (4.0, -0.25) -- (4.65, -0.25);
        \draw [thin] (4.4, -0.3) -- (4.65, -0.3);
        \draw [thin] (4.25, -0.35) -- (4.65, -0.35);
        \draw [thin] (2.8, -0.2) -- (4.65, -0.2);
        \draw [thin] (2.2, -0.15) -- (3.5, -0.15);
        \draw [thin] (3.5, 0.9) -- (3.8, 0.9); 
        \draw [thin] (3.8, -0.15) -- (4.65, -0.15);

        \draw [thin, dotted] (4.25, 1.05) -- (4.25, -0.35);
        \draw [thin, dotted] (4.4, 1.0) -- (4.4, -0.3);
        \draw [thin, dotted] (4.0, 0.95) -- (4.0, -0.25);
        \draw [thin, dotted] (3.5, -0.15) -- (3.5, 0.9);
        \draw [thin, dotted] (3.8, 0.9) -- (3.8, -0.15);
        
        \end{tikzpicture}
        \caption{Complete the paths when the process is intermittently observed using a uniformization sampling algorithm for endpoint conditioned continuous-time Markov chains \citep{hobolth2009simulation}.}
        \label{fig:possiblepathsfull}
    \end{subfigure} %
    \caption{Graphical representation of the algorithm for sampling paths for a latent multistate process from a Markov surrogate. The proposed paths always include the observed path over intervals when the process is continuously observed. The marginal data log-likelihood averages over the complete data log-likelihoods for the collection of completed sample paths.}
    \label{fig:samplingalgorithm}
\end{figure}

\clearpage
\subsection{MCEM Q-function and importance weights}
\label{subsec:importanceweights}
The following is developed at greater depth in \citet{levine2001implementations} but reproduced here for completeness. We drop the dependence on $\bX$ and $\bX_i$ for simplicity. The E-step in the EM algorithm computes the expectation of the complete data log-likelihood over the distribution of the latent process given the observed data. This expectation is referred to as the Q-function. At the $t^{\mr{th}}$ E-step,
\begin{align}
    \label{eqn:mcemqfun}
    Q\left(\btheta,\bthetahat^{(t-1)}\right) 
    &= \int l(\bZ,\bY\mid\btheta)f^\prime\left(\bZ\mid\bY,\bthetahat^{(t-1)}\right)\rmd \bZ \nonumber \\
    &= \int \left(\sum_{i=1}^P l_i(Z_i,\bY_i\mid\btheta)\right) f^\prime\left(\bZ\mid\bY,\bthetahat^{(t-1)}\right)\rmd \bZ \nonumber \\
    &= \sum_{i=1}^P \int l_i(Z_i,\bY_i\mid\btheta)f^\prime\left(\bZ\mid\bY,\bthetahat^{(t-1)}\right)\rmd \bZ \nonumber \\
    &= \sum_{i=1}^P \int l_i(Z_i,\bY_i\mid\btheta)f_i^\prime\left(Z_i\mid\bY_i,\bthetahat^{(t-1)}\right)\rmd Z_i,
\end{align}
where $f^\prime\left(\bZ\mid\bY,\bthetahat^{(t)}\right)=\prod_{i=1}^Pf_i^\prime\left(Z_i\mid\bY_i,\bthetahat^{(t-1)}\right)$ is the density of $\bZ$ conditional on the data, and $f_i^\prime\left(Z_i\mid\bY_i,\bthetahat^{(t-1)}\right)$ the density of $Z_i$ conditional on the data.

We approximate each integral, $\int l_i(Z_i,\bY_i\mid\btheta)f_i^\prime(Z_i\mid\bY_i,\bthetahat^{(t-1)})\rmd Z_i$, in \eqref{eqn:mcemqfun} using importance sampling.
Let $h_i^\prime\left(Z_i\mid\bY_i,\bthetaprime\right)$ be another density with parameters $\bthetaprime$ such that $f_i^\prime$ is absolutely continuous with respect to $h_i^\prime$
. In our case, $f_i^\prime$ is the density of the semi-Markov process and $h_i^\prime$ is the density of the Markov surrogate, both conditional on the data. Rewriting the integral as
\begin{align*}
    \int l_i(Z_i,\bY_i\mid\btheta)f_i^\prime\left(Z_i\mid\bY_i,\bthetahat^{(t-1)}\right)\rmd Z_i &= \int l_i(Z_i,\bY_i\mid\btheta)\frac{f_i^\prime\left(Z_i\mid\bY_i,\bthetahat^{(t-1)}\right)}{h_i^\prime\left(Z_i\mid\bY_i,\bthetaprime\right)}h_i^\prime\left(Z_i\mid\bY_i,\bthetaprime\right)\rmd Z_i,
\end{align*}
we arrive at the approximation based on $M_i$ samples, $Z^{(1)},\dots,Z^{(M_i)}\sim h_i^\prime\left(Z_i\mid\bY_i,\bthetaprime\right)$,
\begin{align}
    \label{eqn:qfunpropweight}
    \int l_i(Z_i,\bY_i\mid\btheta)\frac{f_i^\prime\left(Z_i\mid\bY_i,\bthetahat^{(t-1)}\right)}{h_i^\prime\left(Z_i\mid\bY_i,\bthetaprime\right)}h_i^\prime\left(Z_i\mid\bY_i,\bthetaprime\right)\rmd Z_i \approx \frac{1}{M_i} \sum_{m=1}^{M_i} \nu_{i,m}^{\prime(t-1)}l_i\left(Z_i^{(m)}, \bY_i\mid \btheta\right). 
\end{align}

The importance weights simplify as follows:
\begin{align}
    \label{eqn:mcemweights}
    \nu_{i,m}^{\prime(t-1)} 
    &= \frac{f_i^\prime\left(Z_i^{(m)}\mid\bY_i,\bthetahat^{(t-1)}\right)}{h_i^\prime\left(Z_i^{(m)}\mid\bY_i,\bthetaprime\right)},\nonumber\\
    &= \frac{f_i\left(Z_i^{(m)}\mid\bthetahat^{(t-1)}\right)p_i\left(\bY_i\mid Z_i^{(m)},\bthetahat^{(t-1)}\right)/g_i\left(\bY_i\mid\bthetahat^{(t-1)}\right)}{h_i\left(Z_i^{(m)}\mid\bthetaprime\right)q_i\left(\bY_i\mid Z_i^{(m)},\bthetaprime\right)/r_i\left(\bY_i\mid\bthetaprime\right)},\nonumber\\
    &= \frac{f_i\left(Z_i^{(m)}\mid\bthetahat^{(t-1)}\right)\ind{Z_i^{(m)} \text{ concords with } \bY_i}/g_i\left(\bY_i\mid\bthetahat^{(t-1)}\right)}{h_i\left(Z_i^{(m)}\mid\bthetaprime\right)\ind{Z_i^{(m)} \text{ concords with } \bY_i}/r_i\left(\bY_i\mid\bthetaprime\right)},\nonumber\\
    &\propto \frac{f_i\left(Z_i^{(m)}\mid\bthetahat^{(t-1)}\right)}{h_i\left(Z_i^{(m)}\mid\bthetaprime\right)} = \nu_{i,m}^{(t-1)} .
\end{align}
Note that the indicators that a path concords with the data are always equal to 1 since the paths, $Z_i^{(m)}$, are sampled conditionally on $\bY_i$. The marginal density of the data under the semi-Markov process, $g_i$, cannot be evaluated. This does not pose a problem since we can ignore the ratio $r_i/g_i$, essentially a normalizing constant, in the MCEM algorithm since these terms do not depend on the unknown parameter, $\btheta$, being maximized. Hence, we use the simplified importance weights, $\nu$, instead of the importance weights under the data conditioned Markov proposal, $\nu^\prime$, in the estimate of the integral,
\begin{align}
    \label{eqn:is-estimate-unnormalized}
    l_i(Z_i,\bY_i\mid\btheta)f_i^\prime\left(Z_i\mid\bY_i,\bthetahat^{(t-1)}\right)\rmd Z_i \approx \frac{\sum_{m=1}^{M_i} \nu_{i,m}^{(t-1)}l_i\left(Z_i^{(m)}, \bY_i\mid \bX_i,\btheta\right)}{\sum_{m=1}^{M_i} \nu_{i,m}^{(t-1)}},
\end{align}
which gives the plug-in estimate of the Q-function
\begin{align*}
    \Qanon\left(\btheta,\bthetahat^{(t-1)}\right) &= \sum_{i=1}^P\frac{\sum_{m=1}^{M_i} \nu_{i,m}^{(t-1)}l_i\left(Z_i^{(m)}, \bY_i\mid \bX_i,\btheta\right)}{\sum_{j=1}^{M_i} \nu_{i,j}^{(t-1)}}.
\end{align*}

\subsection{MCEM effective sample size}
\label{subsec:mcemess}

From \citet{liu2001monte}, Section 2.5.3, the effective sample size associated with the importance sampling estimate \eqref{eqn:is-estimate-unnormalized} for person $i$ is, in terms of the self-normalized weights $\tilde{\nu}_{i,m} = \nu_{i,m} / \sum_{j=1}^{M_i}\nu_{i,j}$, $m=1,\dots,M_i$,
\begin{align*}
    \textsf{ESS}_i
    & = \frac{M_i}{1+M_i^2Var_{h^\prime_i}(\tilde{\nu}_{i})} \\
    & = \frac{M_i}{1+M_i^2[E_{h^\prime_i}(\tilde{\nu}_{i}^2) - E_{h^\prime_i}^2(\tilde{\nu}_{i})]} \\
    & = \frac{M_i}{1+M_i^2[E_{h^\prime_i}(\tilde{\nu}_{i}^2) - 1/M_i^2]} \\
    & = \frac{1}{M_i E_{h^\prime_i}(\tilde{\nu}_{i}^2)} \\
    & \approx \frac{1}{\sum_{j=1}^{M_i} \tilde{\nu}_{i,m}^2} = M^\star_i,
\end{align*}
where the third equality holds because $E_{h^\prime}(\tilde{\nu}_i) = 1/M_i$ by construction. If the proposal distribution for the sample paths, $h^\prime_i$ is similar to the target $f^\prime_i$, then $M^\star_i$ will be close to the sample size $M_i$, indicating that the importance sampling procedure is almost as efficient as standard Monte Carlo; while if there is a large discrepancy between $h^\prime_i$ and $f^\prime_i$, then $M^\star_i$ may be much smaller than $M_i$, indicating that a large number of sample paths may be necessary to obtain a stable estimate of the expected log-likelihood \eqref{eqn:is-estimate-unnormalized}.

\subsection{Ascent MCEM decision rules}
\label{subsec:mcemdecisions}

Let $\sigmahat_{ASE}/{M_{Eff}}$ be the estimated asymptotic standard error (ASE) for the change in the Q-function with a target effective sample size of $M_{Eff}$, $\Delta Q_{M_{Eff}}\left(\bthetahat^{(t)},\bthetahat^{(t-1)}\right) = Q_{M_{Eff}}\left(\bthetahat^{(t)},\bthetahat^{(t-1)}\right) - Q_{M_{Eff}}\left(\bthetahat^{(t-1)},\bthetahat^{(t-1)}\right)$, the formula for which is given in equation (14) of \citet{caffo2005ascent}.

\subsubsection{MCEM ascent determination}
\label{subsubsec:mcemascent}
Take $z_\alpha$ such that $\Phi(\alpha) = \Pr(Z\leq z_\alpha) = \alpha$, where $\Phi(\cdot)$ is the cumulative distribution function (CDF) of a standard normal random variable. We accept the maximizer in the $t^{\mr{th}}$ MCEM M-step if there is sufficient evidence that the change in the Q-function is greater than 0, up to Monte Carlo error. The criterion for this is 
\begin{align}
    \label{eqn:ascentruleascend}
    \Delta Q_{M_{Eff}}\left(\bthetahat^{(t)},\bthetahat^{(t-1)}\right) - z_\alpha\sigmahat_{ASE} / \sqrt{M_{Eff}} > 0.
\end{align}
If the lower bound of the asymptotic confidence interval on $\Delta Q_{M_{Eff}}\left(\bthetahat^{(t)},\bthetahat^{(t-1)}\right)$ is negative, we increase the target ESS for each individual by a factor $\kappa>1$. The MCEM E- and M- steps are then repeated with the augmented pool of samples. We iterate between augmenting the pool of samples and the E- and M- steps until (\ref{eqn:ascentruleascend}) is satisfied, at which point we accept the maximizer. 

\subsubsection{MCEM stopping rule}
\label{subsubsec:mcemstopping}
Take $z_\gamma$ such that $\Phi(\gamma) = \Pr(Z\leq z_\gamma) = \gamma$, where $\Phi(\cdot)$ is the cumulative distribution function (CDF) of a standard normal random variable. We terminate after $t^{\mr{th}}$ MCEM M-step if there is sufficient evidence that change in the Q-function is less than a specified tolerance, up to Monte Carlo error. The criterion for this is 
\begin{align}
    \label{eqn:ascentrulestop}
    \Delta Q_{M_{Eff}}\left(\bthetahat^{(t)},\bthetahat^{(t-1)}\right) + z_\gamma\sigmahat_{ASE} / \sqrt{M_{Eff}} < C,
\end{align}
i.e., if the upper bound of the confidence interval for $\Delta Q_{M_{Eff}}\left(\bthetahat^{(t)},\bthetahat^{(t-1)}\right)$ excludes $C.$

\subsubsection{Considerations for MCEM tuning parameters}
\label{subsubsec:tuningpars}
The computation time of our MCEM algorithm is driven by the optimization steps. The algorithm possesses three tuning parameters specified by the user, which impact the number of optimization steps and the cost of these steps: the tolerance $\epsilon>0$, upper bound limit $0<\gamma<1$ and lower bound limit $0<\alpha<1$. Decreasing $\epsilon$ and $\gamma$ make the stopping criterion of the MCEM more stringent, resulting in an estimate of the MLE that is closer to the true MLE at the cost of additional optimization steps. Decreasing $\alpha$ increases the frequency with which the target ESS is increased, resulting in the sampling of additional complete paths for the importance sampling estimate of the Q function in the E-step. A larger number of sample paths result in an importance sampling estimate of the Q function that contains more terms and is therefore more expensive to evaluate in the optimization steps. However, a larger number of sample path decreases the uncertainty in the estimate of the Q function, which may result in an earlier termination of the MCEM algorithm.

\subsection{Observed Fisher information}
\label{subsec:fisherinfo}
Following \citet{louis1982finding}, the observed Fisher information matrix can be expressed as
\begin{align}
    \label{eqn:fisherexpectation}
    I(\btheta\mid\bY) &= -\E\left[\nabla^2_{\btheta} \log f(\bZ;\btheta)\mid\bY\right] - \E\left[(\nabla_{\btheta} \log f(\bZ;\btheta))(\nabla_{\btheta} \log f(\bZ;\btheta))^T\mid\bY\right] + \nonumber\\ & \hspace{2em}\E\left[\nabla_{\btheta} \log f(\bZ;\btheta)\mid\bY\right]\E\left[\nabla_{\btheta} \log f(\bZ;\btheta)^T\mid\bY\right]^T,
\end{align}
where $\nabla_{\btheta} \log f(\bZ;\btheta)$ and $\nabla^2_{\btheta} \log f(\bZ;\btheta)$ are the gradient and hessian of the log-likelihood under the semi-Markov model and the expectations are taken over the distribution of $\bZ$. Given $M$ samples from the Markov surrogate drawn conditionally on the data, $Z^{(1)}_i,\dots,Z^{(M_i)}_i\sim h^\prime_i(Z\mid\bY,\bthetaprime)$ with self-normalized importance weights, $\tilde{\nu}_{i, 1},\dots,\tilde{\nu}_{i, M_i}$, from the final iteration of the MCEM algorithm, the observed fisher information can be approximated at the MLE, $\bthetahat$, via
\begin{align}
    \label{eqn:fisherest}
    \hat{I}(\bthetahat\mid\bY) &= \sum_{i=1}^P M_i\sum_{m=1}^{M_i}\tilde{\nu}_{i,m}\left[-\nabla^2_{\bthetahat}\log f\left(Z^{(m)}_i;\bthetahat\right) - \left(\nabla_{\bthetahat}\log f\left(Z^{(m)}_i;\bthetahat\right)\right)\left(\nabla_{\bthetahat}\log f\left(Z^{(m)}_i;\bthetahat\right)\right)^T\right]\nonumber\\
    & \hspace{2em}+ M_i^2\sum_{m=1}^{M_i}\sum_{n=1}^{M_i}\tilde{\nu}_{i,m}\tilde{\nu}_{i,n}\left[\nabla_{\bthetahat}\log f\left(Z^{(m)}_i;\bthetahat\right)\right]\left[\nabla_{\bthetahat}\log f\left(Z^{(n)}_i;\bthetahat\right)\right]^T,
\end{align}
as derived in \citet{gong2022exact}. In practice, we evaluate gradients and Hessians using forward-mode automatic differentiation as implemented in the \texttt{ForwardDiff.jl} in \texttt{Julia} \citep{revels2016forward}. We then invert the observed Fisher information matrix \eqref{eqn:fisherest} to obtain model-based confidence intervals. 

\subsection{Marginal likelihood estimation}
\label{subsec:marginal-lik}

To simplify notation, we suppress the dependence on the covariate $X_i$.
Under a semi-Markov model, the marginal density of the observed data for participant $i=1,\dots,n$ conditional on $\bthetahat$ is
\begin{align}
\label{eqn:marginal-lik}
    f_i(\bY_i\mid \btheta)
    & = \int f_i(Z_i\mid\btheta) p_i(\bY_i\mid Z_i, \btheta)\rmd Z_i \nonumber \\
    & = \int \frac{f_i(Z_i\mid\btheta) p_i(\bY_i\mid Z_i, \btheta)}{h^\prime_i(Z_i\mid \bY_i, \bthetaprime)} h^\prime_i(Z_i\mid \bY_i, \bthetaprime)\rmd Z_i \nonumber \\
    & = \int \frac{f_i(Z_i\mid\btheta) p_i(\bY_i\mid Z_i, \btheta)}{h_i\left(Z_i\mid\bthetaprime\right)q_i\left(\bY_i\mid Z_i,\bthetaprime\right)/r_i\left(\bY_i\mid\bthetaprime\right)} h^\prime_i(Z_i\mid \bY_i, \bthetaprime)\rmd Z_i \nonumber \\
    & = r_i\left(\bY_i\mid\bthetaprime\right)\int \frac{f_i(Z_i\mid\btheta) \ind{Z_i \textsf{ concords with }\bY_i}}{h_i\left(Z_i\mid\bthetaprime\right)\ind{Z_i \textsf{ concords with }\bY_i}} h^\prime_i(Z_i\mid \bY_i, \bthetaprime)\rmd Z_i \nonumber \\
    & = r_i\left(\bY_i\mid\bthetaprime\right)\int \frac{f_i(Z_i\mid\btheta)}{h_i\left(Z_i\mid\bthetaprime\right)} h^\prime_i(Z_i\mid \bY_i, \bthetaprime)\rmd Z_i
\end{align}
where, under a Markov surrogate model, the marginal distribution of the data, $r_i\left(\bY_i\mid\bthetaprime\right)$, can be evaluated.

Equation \eqref{eqn:marginal-lik} suggests the use of the Monte Carlo estimator
\begin{equation}  \label{eq:marginal-lik-est}
    \hat{f}_{i}\left(\bY_i\mid \btheta^{(t)}\right) = \frac{r_i\left(\bY_i\mid\bthetaprime\right)}{M_i}\sum_{m=1}^{M_i} \nu_{i,m}^{(t)} = r_i\left(\bY_i\mid\bthetaprime\right) \bar{\nu}_{i}^{(t)}
\end{equation}
for paths $Z^{(m)}_i$ sampled independently from the Markov model with parameters $\bthetaprime$, where $\bar{\nu}_{i}^{(t)}$ is the empirical average of the sampling weights.

The plug-in estimate of the marginal density of the observed data for the entire sample is then
\begin{equation}
    \hat{f}\left(\bY\mid \btheta^{(t)}\right) = \prod_{i=1}^P \hat{f}_{i}\left(\bY_i\mid \btheta^{(t)}\right) = r(\bY \mid \bthetaprime) \prod_{i=1}^P \bar{\nu}_{i}^{(t)},
\end{equation}
because $r(\bY \mid \bthetaprime)=\prod_{i=1}^P r_i(\bY_i \mid \bthetaprime)$. Note that the factor $r(\bY \mid \bthetaprime)$ is independent of the semi-Markov model and can, therefore, be considered as a normalizing constant when comparing models.

The variance of the estimator~\eqref{eq:marginal-lik-est} is
\begin{equation*} 
    \Var\left( \hat{f}_{i}\left(\bY_i\mid \btheta^{(t)}\right) \right) = (r_i(\bY_i \mid \bthetaprime))^2 \Var\left(\bar{\nu}_{i}^{(t)}\right),
\end{equation*}
which we estimate with
\begin{equation*}
    \hat{\Var}\left( \hat{f}_{i}\left(\bY_i\mid \btheta^{(t)}\right) \right) = (r_i(\bY_i \mid \bthetaprime))^2 \hat{\Var}\left(\bar{\nu}_{i}^{(t)}\right),
\end{equation*}
with the unbiased estimate
\begin{equation*}
    \hat{\Var}\left(\bar{\nu}_{i}^{(t)}\right) = \frac{1}{M_i} \frac{1}{M_i-1} \sum_{m=1}^{M_i} \left(\nu_{i,m}^{(t)} - \bar{\nu}_{i}^{(t)}\right)^2.
\end{equation*}
Increasing the number of sampled paths, $M_i$, reduces $\Var\left(\bar{\nu}_{i}^{(t)}\right)$.

When comparing multistate semi-Markov models using the information criteria AIC and BIC, one needs an estimate of the log marginal likelihood at the MLE found at the last step, $t^\star$, of the MCEM algorithm. We propose the estimate
\begin{equation}
\label{eqn:log-marginal-lik}
    \log \hat{f}\left(\bY\mid \btheta^{(t^\star)}\right) = \log r(\bY \mid \bthetaprime) + \sum_{i=1}^P \log \bar{\nu}_{i}^{(t^\star)}.
\end{equation}
By the delta method, we have for large $M_i$
\begin{equation*}
    \Var\left(\log \bar{\nu}_{i}^{(t^\star)}\right) \approx \frac{\Var\left(\bar{\nu}_{i}^{(t^\star)}\right)}{\left(\bar{\nu}_{i}^{(t^\star)}\right)^2}.
\end{equation*}
We therefore estimate the variance of the estimate of the log marginal likelihood \eqref{eqn:log-marginal-lik} with
\begin{equation}  \label{eq:var-log-marginal-lik-est}
    \hat{\Var}\left(\log \hat{f}\left(\bY\mid \btheta^{(t^\star)}\right) \right) = \sum_{i=1}^P \frac{\Var\left(\bar{\nu}_{i}^{(t^\star)}\right)}{\left(\bar{\nu}_{i}^{(t^\star)}\right)^2}.
\end{equation}
The variance estimate~\eqref{eq:var-log-marginal-lik-est} can be used to determine whether the estimated AIC of two models are significantly different.

\newpage
\section{Technical Appendix: Details of the simulation study for assessment of intercurrent events with an illness-death model in Section \ref{subsec:illnessdeath}}
\label{sec:illnessdeathsupp}

\subsection{Simulation model}
\label{subsec:illnessdeath_simulation_model}

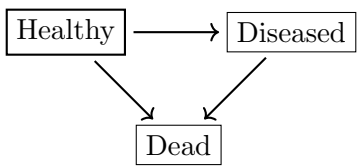
\begin{figure}[htbp]
    \centering
    \begin{tikzpicture}[scale=1, every node/.style={scale=1}, every text node part/.style={align=center}]
        \draw (0.0, 0.0) node[align=center, draw,rectangle,thick] (1) {Healthy};
            \draw (3.0, 0.0) node[align=center, draw, rectangle] (2) {Diseased};
            \draw (1.5, -1.5) node[align=center, draw, rectangle] (3) {Dead};
        
            \draw [thick,->,shorten >=1mm,shorten <=1mm] (1) to (2) node {};
            \draw [thick,->,shorten >=1mm,shorten <=1mm] (2) to (3) node {};
            \draw [thick,->,shorten >=1mm,shorten <=1mm] (1) to (3) node {};
    \end{tikzpicture}
    \caption{Direct transitions between states in a progressive illness-death model.}
    \label{fig:illnessdeath_diagram}    
\end{figure}

We simulate clinical histories for 250 study participants from a progressive illness-death model, diagrammed in Figure \ref{fig:illnessdeath_diagram}. All participants begin follow-up upon entrance into the healthy state. This might be, for instance, the time at which a cancer has been considered to have gone into remission. Participants are followed for one year and their clinical status is observed at either monthly increments or every three months. We add some variation around the scheduled assessment time to better emulate a real-world study. The random visit times are drawn from a beta(1.5, 1.5) distribution centered around the scheduled time and scaled to span the midpoints between scheduled assessments, e.g., using the verbiage of random number generation in \texttt{R}, an individual's visit time for month 2 would be drawn as 2 + (rbeta(1, 1.5, 1.5) - 0.5). The time of enrollment and end of follow-up were 0 and 1 year for all participants.  

All three transitions had Weibull baseline intensities with parameters given in Table \ref{tab:illnessdeath_pars}. The true values for the quantities of interest estimated in the simulation in Section \ref{subsec:illnessdeath} are given in Table \ref{tab:illnessdeath_truth}.

\begin{table}[htbp]
    \centering
    \begin{tabular}{lcl}
    \toprule
        \textbf{Transition} & \textbf{Baseline intensity} & \textbf{Parameters}\\
        \cmidrule(lr){1-1}\cmidrule(lr){2-2}\cmidrule(lr){3-3}Healthy$\longrightarrow$Diseased& Weibull & $\lambda_0 = 1.5$, $\kappa = 1.25$ \\
        Healthy$\longrightarrow$Dead& Weibull & $\lambda_0 = 1$, $\kappa = 1.25$ \\
        Diseased$\longrightarrow$Dead& Weibull & $\lambda_0 = 2$, $\kappa = 1.25$ \\\bottomrule
    \end{tabular}
    \caption{Parameterization and parameter values for transition intensities in the simulation model depicted in Figure \ref{fig:illnessdeath_diagram}. Weibull log-intensities are parameterized as $\log(\lambda_i(t, Z_i)) = \log(\lambda_0) + \log(\alpha) + t(\kappa - 1)$, where $\lambda_0 > 0$ and $\kappa > 0$.}
    \label{tab:illnessdeath_pars}
\end{table}

\begin{table}[htbp]
    \centering
    \begin{tabular}{lc}
         \toprule 
         \textbf{Quantity}& \textbf{Truth} \\
         \cmidrule(lr){1-1}\cmidrule(lr){2-2} Pr(Recurrence-free survival)& 0.082 \\
         Pr(Recurrence) & 0.551 \\
         Pr(Death w/ recur.) & 0.349 \\
         Pr(Death w/o recur.) & 0.367 \\
         RM RFST & 0.423 \\
         Time to recur. $\mid$ recur. & 0.371 \\
         RM time to recur. or EoF & 0.654 \\
         RM time to death $\mid$ recur. & 0.378 \\\bottomrule
    \end{tabular}
    \caption{Ground truth for key quantities for the illness-death simulation in Section \ref{subsec:illnessdeath}. Time is given in years. RM RFST is the restricted mean (RM) recurrence-free survival time, which is the minimum of the time to recurrence, death, or end of follow-up (EoF). Time to recurrence conditional on recurrence is the time to recurrence among individuals whose disease recurred. RM time to recurrence or EoF is the minimum of the time to recurrence or EoF, treating participants who die prior to recurrence as having recurrence time equal to the duration of follow-up. RM time to death conditional on recurrence is the minimum of the time to death or EoF starting at disease recurrence.}
    \label{tab:illnessdeath_truth}
\end{table}

\subsection{Models for inference}
\label{subsec:illnessdeath_infmodels}

\subsubsection{Multistate models}
\label{subsubsec:illnessdeath_msms}

Four illness-death models were fit using the parameterizations given in Table \ref{tab:illnnessdeath_parameterization}. For each model, all transition intensities were assigned the same functional form. Two of the models were semi-parametric and used B-splines to approximate the baseline intensity for each transition. The B-spline bases were computed using the \texttt{BSplineKit.jl} package in \texttt{Julia} \citep{bsplinekit}. In the simple case, we used degree 1 B-splines, i.e., linear splines, with a single interior knot set at the median observed transition time for each possible event. In the more complicated case, we used natural cubic splines with interior knots at the 1/3 and 2/3 quantiles of the transition times for each event. Note that for the transition from ill to dead, we computed the quantiles for time from recurrence to death by taking the left endpoint of the interval in which disease was known to have recurred as the time of illness onset. In both spline models, the right boundary knot was set to the greatest observed transition time, or the maximum of the right endpoints of the intervals in which recurrence was known to have occurred in the case of the healthy to ill transition. We used a flat extrapolation of the transition intensity beyond the right spline boundary.  

\begin{table}[htbp]
    \centering
    \def\arraystretch{1.5}
    \begin{tabular}{lll}
    \toprule
        Baseline intensity & Parameterization & Constraints\\
        \cmidrule(lr){1-1} \cmidrule(lr){2-2} \cmidrule(lr){3-3} Exponential& $\log(\lambda_{ij}(t)) = \log(\lambda_{ij})$ & $\lambda_{ij} > 0$\\
        Weibull & $\log(\lambda_{ij}(t)) = \log(\lambda_{ij}) + \log(\alpha) + t(\kappa - 1) $ & $\lambda_{ij} > 0$ and $\kappa_{ij} > 0$\\
        B-spline & $\log(\lambda_{ij}(t)) = \log\left(\sum_{\ell = 1}^L\gamma_\ell B_\ell\left(t;\boldsymbol{\xi},\kappa\right)\right)$ & $\boldsymbol{\gamma} > 0$\\
        \bottomrule
    \end{tabular}
    \caption{Parameterization of transition intensities for illness-death models used in Section \ref{subsec:illnessdeath}. The log baseline intensity from state $i$ to state $j$ is denoted $\log(\lambda_{ij}(t))$.  $B_\ell(t;\boldsymbol{\xi},\kappa)$ is the $\ell-$th basis function for an B-spline with interior knots $\boldsymbol{\xi}$ and degree $\kappa$, evaluated at time $t$.}
    \label{tab:illnnessdeath_parameterization}
\end{table}

\clearpage
\section{Technical Appendix: Details of the simulation study for treatment efficacy and natural history in Section \ref{subsec:sim2regen}}
\label{sec:regensimsupp}

\subsection{Simulation model}
\label{subsec:regen_fullmodel}

We assume that REGEN-2069 participants are na\"{\i}ve at randomization, that is, uninfected and at high risk of exposure to SARS-CoV-2 due to occurrence of COVID-19 within their household within 96 hours of study enrollment. Study participants are confirmed SARS-CoV-2 negative by RT-qPCR at enrollment, meaning that they do not have detectable virus. There are two states of RT-qPCR status that follow infection -- confirmed active SARS-CoV-2 infection as detected by RT-qPCR, denoted PCR+, and clearance of SARS-CoV-2 infection as indicated by a negative assay following confirmed infection, which is denoted PCR$-$. All infected participants are assumed to have detectable virus, though the duration of RT-qPCR positivity may be short. Hence, weekly testing may not be sufficiently dense to detect some cases. Symptomatic COVID-19 is confirmed by RT-qPCR at symptom onset. We include in the model a state for whether a participant has been symptomatically infected. Finally, we define seroconversion by whether a participant has detectable antibodies to the SARS-CoV-2 nucleocapsid. 

The states in the simulation model are enumerated in Table \ref{tab:regenstates}. We deduce the set of allowable transitions depicted in Figure \ref{subfig:regendiagram} on the basis of the assumptions in Table \ref{tab:regenassump}, yielding 9 possible transitions. The parameterizations of the transition intensities and true parameter values are shown in Table \ref{tab:regensim_pars}. The ground truths for key functionals of interest under the true simulation parameters is shown in Table \ref{tab:regensim_truth}. 

\begin{table}[htbp]
    \centering
    \begin{tabular}{cccccc}
            \toprule
             & \multicolumn{5}{c}{State definition}\\
            \cmidrule(lr){2-6} State & Na\"{\i}ve & PCR+ & PCR$-$- & \shortstack{Ever \\ Sympt.} & Sero.+\\
            \cmidrule(lr){1-1}\cmidrule(lr){2-6}1 & + & -- & -- & -- &-- \\
            2 & -- & + & -- & -- &-- \\
            3 & -- & + & -- & -- & + \\
            4 & -- & -- & + & -- &-- \\
            5 & -- & -- & + & -- & + \\
            6 & -- & + & -- & + & -- \\
            7 & -- & + & -- & + & + \\
            8 & -- & -- & + & + & -- \\
            9 & -- & -- & + & + & + \\
            \bottomrule
        \end{tabular}    
        \caption{State definitions for clinical and immunological progression in the REGEN-2069 trial. Na\"{\i}ve is uninfected at randomization, PCR+ is detectable by RT-qPCR, PCR$-$- is previously RT-qPCR+ and no longer detectable by RT-qPCR, Ever Sympt. is ever symptomatic, Sero.+ is detectable antibody to the SARS-CoV-2 nucleocapsid.}
        \label{tab:regenstates}
\end{table}

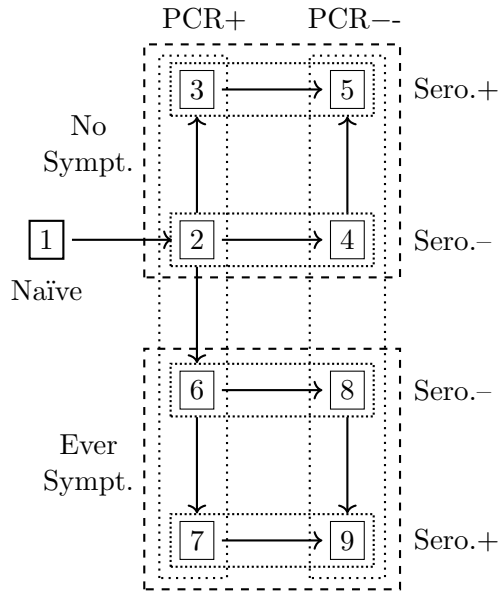
\begin{figure}\centering
    \begin{tikzpicture}[scale=1, every node/.style={scale=1}, every text node part/.style={align=center}]
        \draw (-01, 0.0) node[align=center, draw,rectangle,thick] (1) {1};
            \draw (1.0, 0.0) node[align=center, draw, rectangle] (2) {2};
            \draw (1.0, 2) node[align=center, draw, rectangle] (3) {3};
            \draw (3.0, 0.0) node[align=center, draw, rectangle] (4) {4};
            \draw (3.0, 2) node[align=center, draw, rectangle] (5) {5};
            \draw (1.0, -2) node[align=center, draw, rectangle] (6) {6};
            \draw (1.0, -4) node[align=center, draw, rectangle] (7) {7};
            \draw (3.0, -2) node[align=center, draw, rectangle] (8) {8};
            \draw (3.0, -4) node[align=center, draw, rectangle] (9) {9};
        
            \draw [thick,->,shorten >=1mm,shorten <=1mm] (1) to (2) node[midway,above] {};
            \draw [thick,->,shorten >=1mm,shorten <=1mm] (2) to (3) node[midway,above] {};
            \draw [thick,->,shorten >=1mm,shorten <=1mm] (2) to (4) node[midway,above] {};
            \draw [thick,->,shorten >=1mm,shorten <=1mm] (3) to (5) node[midway,above] {};
            \draw [thick,->,shorten >=1mm,shorten <=1mm] (4) to (5) node[midway,above] {};

            \draw [thick,->,shorten >=1mm,shorten <=1mm] (2) to (6) node[midway,above] {};
            \draw [thick,->,shorten >=1mm,shorten <=1mm] (6) to (7) node[midway,above] {};
            \draw [thick,->,shorten >=1mm,shorten <=1mm] (7) to (9) node[midway,above] {};
            \draw [thick,->,shorten >=1mm,shorten <=1mm] (6) to (8) node[midway,above] {};
            \draw [thick,->,shorten >=1mm,shorten <=1mm] (8) to (9) node[midway,above] {};

            \draw[thick,dotted] (0.5,-4.5) rectangle (1.4,2.45);
            \draw[thick,dotted] (2.5,-4.5) rectangle (3.5,2.45);
            \draw[thick,dashed] (0.3,-0.5) rectangle (3.7,2.6);
            \draw[thick,dashed] (0.3,-4.65) rectangle (3.7,-1.45);
            \draw[thick,densely dotted] (0.65,-4.35) rectangle (3.35, -3.65);
            \draw[thick,densely dotted] (0.65,-2.35) rectangle (3.35, -1.65);
            \draw[thick,densely dotted] (0.65, 1.65) rectangle (3.35, 2.35);
            \draw[thick,densely dotted] (0.65, -0.35) rectangle (3.35, 0.35);
            
            \node[anchor=north] at (-1,-0.4) {Na\"{\i}ve};
            \node[anchor=east] at (0.3, 1.25) {No \\Sympt.};
            \node[anchor=east] at (0.3,-3) {Ever\\Sympt.};
            \node[anchor=west] at (3.75, 2) {Sero.+};
            \node[anchor=west] at (3.75,-4) {Sero.+};
            \node[anchor=west] at (3.75, 0.0) {Sero.--};
            \node[anchor=west] at (3.75,-2) {Sero.--};
            \node[anchor=south] at (1.1,2.65) {PCR+};
            \node[anchor=south] at (3.1,2.65) {PCR$-$-};
    \end{tikzpicture}
    \caption{Direct transitions between states in the model for clinical and immunological progression from which participant histories were simulated in Section \ref{subsec:sim2regen}.}
    \label{fig:regensimdiagram}    
\end{figure}

\begin{table}[htbp]
    \centering
    \begin{tabular}{ccl}
    \toprule
        Transition & Baseline intensity & Parameters\\
        \cmidrule(lr){1-1}\cmidrule(lr){2-2}\cmidrule(lr){3-3}$1\longrightarrow2$ & Weibull & $\lambda_0 = 0.6$, $\kappa = 0.7$, $\beta = \log(0.33)$ \\
        $2\longrightarrow3$ & Exponential & $\lambda_0 = 0.2$, $\beta = \log(0.5)$ \\
        $2\longrightarrow4$ & Exponential & $\lambda_0 = 0.3$, $\beta = 0$ \\
        $3\longrightarrow4$ & Exponential & $\lambda_0 = 0.3$, $\beta = 0$ \\
        $4\longrightarrow5$ & Exponential & $\lambda_0 = 0.25$, $\beta = \log(0.5)$\\
        $2\longrightarrow6$ & Weibull & $\lambda_0 = 1$, $\kappa = 1.5$, $\beta = \log(0.4)$\\
        $6\longrightarrow7$ & Exponential & $\lambda_0 = 1$, $\beta = \log(0.5)$\\
        $6\longrightarrow8$ & Exponential & $\lambda_0 = 0.5$, $\beta = 0$\\
        $7\longrightarrow9$ & Exponential & $\lambda_0 = 0.5$, $\beta = 0$\\
        $8\longrightarrow9$ & Exponential & $\lambda_0 = 1$, $\beta = \log(0.5)$\\\bottomrule
    \end{tabular}
    \caption{Parameterization and parameter values for transition intensities in the simulation model depicted in Figure \ref{fig:regensimdiagram}. Exponential log-intensities are parameterized as $\log(\lambda_i(t, Z_i)) = \log(\lambda_0) + \beta Z_i$, where $\lambda_0>0$ and $Z_i$ is the treatment assignment indicator for participant $i$ with value 0 if $i$ is assigned to placebo and 1 if assigned to mAb. Weibull log-intensities are parameterized as $\log(\lambda_i(t, Z_i)) = \log(\lambda_0) + \log(\alpha) + t(\kappa - 1) + \beta Z_i$, where $\lambda_0 > 0$ and $\kappa > 0$.}
    \label{tab:regensim_pars}
\end{table}

\clearpage

\begin{table}[htbp]
  \centering
  \begin{tabular}{rc}
    \toprule
    \textbf{Quantity} & \textbf{Truth}\\
    \cmidrule(lr){1-1}\cmidrule(lr){2-2}Pr(Infec. $\mid$ mAb) & 0.407 \\
    Pr(Infec. $\mid$ Plac.) & 0.795 \\
    \cmidrule(lr){1-2}Pr(Sympt. $\mid$ mAb) & 0.124 \\
    Pr(Sympt. $\mid$ Plac.) & 0.430 \\
    \cmidrule(lr){1-2}Pr(Asympt. $\mid$ mAb) & 0.283 \\
    Pr(Asympt. $\mid$ Plac.) & 0.365 \\
    \cmidrule(lr){1-2}Pr(Sympt. $\mid$ Infec., mAb) & 0.304 \\
    Pr(Sympt. $\mid$ Infec., Plac.) & 0.54 \\
    Pr(Asympt. $\mid$ Infec., mAb) & 0.696 \\
    Pr(Asympt. $\mid$ Infec., Plac.) & 0.460 \\
    \cmidrule(lr){1-2}Pr(N+ $\mid$ Infec., mAb) & 0.378 \\
    Pr(N+ $\mid$ Infec., Plac.) & 0.715 \\
    Pr(N+ $\mid$ sympt., mAb) & 0.572 \\
    Pr(N+ $\mid$ sympt., Plac.) & 0.837 \\
    Pr(N+ $\mid$ asympt., mAb) & 0.293 \\
    Pr(N+ $\mid$ asympt., Plac.) & 0.570 \\
    \cmidrule(lr){1-2}RM TI $\mid$ mAb & 2.97 \\
    RM TI $\mid$ Plac. & 1.730 \\
    \cmidrule(lr){1-2}RM PCR+ $\mid$ mAb & 1.95 \\
    RM PCR+ $\mid$ Plac. & 1.52 \\
    Pr(Detected $\mid$ mAb) & 0.843 \\
    Pr(Detected $\mid$ Plac.) & 0.646 \\
    \cmidrule(lr){1-2}PE for infec. & 0.488 \\
    PE for sympt. infec. & 0.712 \\
    PE for asympt. infec. & 0.224 \\
    \cmidrule(lr){1-2} RR for sympt. $\mid$ infec. & 0.562 \\
    RR for N+ $\mid$ infec. & 0.528 \\
    RR for N+ $\mid$ sympt. & 0.684 \\
    RR for N+ $\mid$ asympt. & 0.513 \\\bottomrule
  \end{tabular}
  \caption{Ground truth for functionals of interest in simulated trials of a monoclonal antibody (mAb) prophylaxis. We abbreviate as follows: Pr = Probability, RMIFT = restricted mean infection free time, RM PCR+ = restricted mean duration of PCR+, PE = protective efficacy, RR = relative risk, N+ = seropositive. RM TI is the arithmetic mean over individuals of the the predicted time to infection or maximum follow-up for each individual, whichever is less. RM PCR+ is the arithmetic mean over individuals of the time during which they are detectable by PCR or the maximum folloowup for each individual, whichever is less. RR is the ratio of probabilities of the specified event, mAb divided by placebo. PE is 1 - RR, i.e., the reduction in relative risk. Ground truths are obtained by Monte Carlo simulation using simulated 100,000 trials.}
  \label{tab:regensim_truth}
\end{table}

\clearpage
\subsection{Parameterization of transition intensities}
\label{subsec:regensim_inferencemods}
\begin{table}[htbp]
    \centering
    \begin{tabular}{ccc}
    \toprule
        & \multicolumn{2}{c}{Transition intensity}\\
        \cmidrule(lr){2-3}Transition & Fully parametric & Semi-parametric\\
        \cmidrule(lr){1-1}\cmidrule(lr){2-2}\cmidrule(lr){3-3}$1\longrightarrow2$ & Weibull & B-spline, $\kappa = 1,\ \xi_1 = 5$ days \\
        $2\longrightarrow3$ & Exponential & B-spline, $\kappa = 1$, $\xi_1 = 7$ days \\
        $2\longrightarrow4$ & Weibull & B-spline, $\kappa = 1$, $\xi_1 = 7$ days \\
        $4\longrightarrow5$ & Exponential & B-spline, $\kappa = 1$, $\xi_1 = 7$ days \\
        \bottomrule
    \end{tabular}
    \caption{Parameterization of transition intensities for the five-state inferential models used in Section \ref{subsec:sim2regen} that are depicted in Figure \ref{subfig:regendiagram}. Exponential log-intensities are parameterized as $\log(\lambda_i(t, Z_i)) = \log(\lambda_0) + \beta Z_i$, where $\lambda_0>0$ and $Z_i$ is the treatment assignment indicator for participant $i$ with value 0 if $i$ is assigned to placebo and 1 if assigned to mAb. Weibull log-intensities are parameterized as $\log(\lambda_i(t, Z_i)) = \log(\lambda_0) + \log(\alpha) + t(\kappa - 1) + \beta Z_i$, where $\lambda_0 > 0$ and $\kappa > 0$. B-spline transition intensities are parameterized as $\log(\lambda_i(t,Z_i)) = \log\left(\sum_{\ell = 1}^L\gamma_\ell B_\ell\left(t;\boldsymbol{\xi},\kappa\right)\right) + \beta Z_i$, where $\boldsymbol{\gamma}>0$ are the B-spline coefficients and $B_\ell(t;\boldsymbol{\xi},\kappa)$ is the $\ell-$th basis function for an B-spline with interior knots $\boldsymbol{\xi}$ and degree $\kappa$ evaluated at time $t$.}
    \label{tab:regensim_paraterization}
\end{table}

\newpage
\subsection{Crude estimation by tabulation of observable quantities}
\label{subsubsec:crudeests}

Notation for counts, subscripts denote assignment to placebo (P) or treatment (T):

\begin{itemize}
    \item $C_P,\ C_T$: counts of participants per arm,
    \item $I_P,\ I_T$: counts of infected participants by PCR+, symptoms, or seropositivity,
    \item $S_P,\ S_T$: counts of symptomatically infected participants,
    \item $A_P = I_P - S_P,\ A_T = I_T - S_T$: counts of asymptomatically infected participants,
    \item $N_P,\ N_T$: counts of seropositive participants,
    \item $N_{P,S},\ N_{T,S}$: counts of seropositive symptomatic participants,
    \item $N_{P,A} = N_P - N_{P,S},\ N_{T,A} = N_T - N_{T,S}$: counts of seropositive asymptomatic participants. 
\end{itemize}

Crude estimators of key quantities of interest are shown below. Confidence intervals for binomial proportions are computed using the Wilson score method in the \texttt{binom} package in \texttt{R} \citep{dorairaj2022binom}. Confidence intervals for ratios of bionomial proportions are computed using the Koopman method in the \texttt{DescTools} package in \texttt{R} \citep{signorell2023DescTools}.

\begin{itemize}
    \item $\mr{Pr(Infec. }\mid \mr{Plac.)} = \nicefrac{I_P}{C_P}$, $\mr{Pr(Infec. }\mid \mr{Trt.)} = \nicefrac{I_T}{C_T}$,
    \item $\mr{Pr(Sympt. }\mid \mr{Plac.)} = \nicefrac{S_P}{C_P}$, $\mr{Pr(Sympt. }\mid \mr{Trt.)} = \nicefrac{S_T}{C_T}$,
    \item $\mr{Pr(Asympt. }\mid \mr{Plac.)} = \nicefrac{A_P}{C_P}$, $\mr{Pr(Asympt. }\mid \mr{Trt.)} = \nicefrac{A_T}{C_T}$,
    \item $\mr{Pr(Sympt. }\mid \mr{Infec.,\ Plac.)} = \nicefrac{S_P}{I_P}$, $\mr{Pr(Sympt. }\mid \mr{Infec.,\ Trt.)} = \nicefrac{S_T}{I_T}$,
    \item $\mr{Pr(Asympt. }\mid \mr{Infec.,\ Plac.)} = \nicefrac{A_P}{I_P}$, $\mr{Pr(Asympt. }\mid \mr{Infec.,\ Trt.)} = \nicefrac{A_T}{I_T}$,
    \item $\mr{Pr(Seropos. }\mid \mr{Infec.,\ Plac.)} = \nicefrac{N_P}{I_P}$, $\mr{Pr(Seropos. }\mid \mr{Infec.,\ Trt.)} = \nicefrac{N_T}{I_T}$,
    \item $\mr{Pr(Seropos. }\mid \mr{Sympt.,\ Plac.)} = \nicefrac{N_P}{S_P}$, $\mr{Pr(Seropos. }\mid \mr{Sympt.,\ Trt.)} = \nicefrac{N_T}{S_T}$,
    \item $\mr{Pr(Seropos. }\mid \mr{Asympt.,\ Plac.)} = \nicefrac{N_P}{A_P}$, $\mr{Pr(Seropos. }\mid \mr{Asympt.,\ Trt.)} = \nicefrac{N_T}{A_T}$,
    \item $\mr{PE(Infec.)} = 1 - \nicefrac{\mr{\Pr(Infec.\mid Trt.)}}{\mr{\Pr(Infec.\mid Plac.)}}$, 
    \item $\mr{PE(Sympt.)} = 1 - \nicefrac{\mr{\Pr(Sympt.\mid Trt.)}}{\mr{\Pr(Sympt.\mid Plac.)}}$,
    \item $\mr{PE(Asympt.)} = 1 - \nicefrac{\mr{\Pr(Asympt.\mid Trt.)}}{\mr{\Pr(Asympt.\mid Plac.)}}$,
    \item $\mr{RR(Sympt. \mid Infec.)} = \nicefrac{\mr{\Pr(Sympt. \mid Infec.,\ Trt.)}}{\mr{\Pr(Sympt.\mid Infec.\ Plac.)}}$,
    \item $\mr{RR(Seropos. \mid Infec.)} = \nicefrac{\mr{\Pr(Seropos. \mid Infec.,\ Trt.)}}{\mr{\Pr(Seropos.\mid Infec.\ Plac.)}}$,
    \item $\mr{RR(Seropos. \mid Sympt.)} = \nicefrac{\mr{\Pr(Seropos. \mid Sympt.,\ Trt.)}}{\mr{\Pr(Seropos.\mid Sympt.\ Plac.)}}$,
    \item $\mr{RR(Seropos. \mid Asympt.)} = \nicefrac{\mr{\Pr(Seropos. \mid Asympt.,\ Trt.)}}{\mr{\Pr(Seropos.\mid Asympt.\ Plac.)}}$.
\end{itemize}

\clearpage
\subsection{Additional results}
\label{subsec:regensim_suppresults}
\begin{table}[htbp]
    \centering\footnotesize
    \begin{tabular}{rrrr}
          \toprule\multicolumn{4}{c}{Complete results using the 9 state simulation model}\\
          \cmidrule(lr){1-4} & \textbf{Bias} & \textbf{Covg.} & \textbf{CIW} \\
          \cmidrule(lr){2-4}Pr(Infec. $\mid$ Plac.) & 0.01 & 0.96 & 0.07 \\
          Pr(Infec. $\mid$ mAb) & 0.0 & 0.95 & 0.17 \\
          \cmidrule(lr){1-4}Pr(Sympt. $\mid$ Plac.) & 0.01 & 0.96 & 0.16 \\
          Pr(Sympt. $\mid$ mAb) & 0.0 & 0.96 & 0.37 \\
          \cmidrule(lr){1-4}Pr(Asympt. $\mid$ Plac.) & 0.0 & 0.96 & 0.18 \\
          Pr(Asympt. $\mid$ mAb) & 0.0 & 0.95 & 0.22 \\
          \cmidrule(lr){1-4}Pr(Sympt. $\mid$ Infec., Plac.) & 0.0 & 0.96 & 0.14 \\
          Pr(Sympt. $\mid$ Infec., mAb) & 0.0 & 0.96 & 0.32 \\
          Pr(Asympt. $\mid$ Infec., Plac.) & 0.0 & 0.96 & 0.16 \\
          Pr(Asympt. $\mid$ Infec., mAb) & 0.0 & 0.96 & 0.14 \\
          \cmidrule(lr){1-4}Pr(N+ $\mid$ Infec., Plac.) & 0.0 & 0.96 & 0.08 \\
          Pr(N+ $\mid$ Infec., mAb) & 0.0 & 0.96 & 0.26 \\
          Pr(N+ $\mid$ sympt., Plac.) & 0.0 & 0.95 & 0.07 \\
          Pr(N+ $\mid$ sympt., mAb) & -0.01 & 0.96 & 0.3 \\
          Pr(N+ $\mid$ asympt., Plac.) & 0.0 & 0.95 & 0.18 \\
          Pr(N+ $\mid$ asympt., mAb) & 0.0 & 0.96 & 0.4 \\
          \cmidrule(lr){1-4}RM TI $\mid$ Plac. & 0.0 & 0.95 & 0.12 \\
          RM TI $\mid$ mAb & 0.0 & 0.95 & 0.07 \\
          \cmidrule(lr){1-4}RM PCR+ $\mid$ Plac. & 0.0 & 0.95 & 0.11 \\
          RM PCR+ $\mid$ mAb & -0.01 & 0.96 & 0.18 \\
          \cmidrule(lr){1-4}Pr(Detected $\mid$ Plac.) & 0.0 & 0.95 & 0.07 \\
          Pr(Detected $\mid$ mAb) & 0.0 & 0.96 & 0.11 \\
          \cmidrule(lr){1-4}PE for infec. & 0.0 & 0.96 & 0.19 \\
          PE for sympt. infec. & 0.0 & 0.96 & 0.16 \\
          PE for asympt. infec. & 0.01 & 0.95 & 0.97 \\
          \cmidrule(lr){1-4}RR for sympt. $\mid$ infec. & 0.0 & 0.96 & 0.35 \\
          RR for N+ $\mid$ infec. & 0.0 & 0.96 & 0.28 \\
          RR for N+ $\mid$ sympt. & 0.0 & 0.96 & 0.3 \\
          RR for N+ $\mid$ asympt. & 0.0 & 0.96 & 0.44 \\
          \bottomrule
    \end{tabular}
    \caption{Bias, coverage, and 95\% confidence interval width of estimates of key quantities obtained by fitting the 9 state simulation model described in Supplementary Appendix \ref{subsec:regen_fullmodel} to hypothetical panel data where PCR and serology were assessed weekly. RM TI is the restricted mean time to infection, RM PCR+ is the restricted mean duration of PCR positivity, RR is relative risk, and PE is protective efficacy, calculated as $PE = 1 - RR$.}
    \label{tab:regensim_fullmod_res}
\end{table}

\clearpage

\begin{table}[htbp]
    \centering
    \begin{tabular}{rrrr}
          \toprule\multicolumn{4}{c}{Complete results using a fully parametric 5 state model}\\
          \cmidrule(lr){1-4} & \textbf{Bias} & \textbf{Covg.} & \textbf{CIW} \\
          \cmidrule(lr){2-4}Pr(Infec. $\mid$ Plac.) & 0.0 & 0.95 & 0.07 \\
              Pr(Infec. $\mid$ mAb) & 0.0 & 0.93 & 0.16 \\
              \cmidrule(lr){1-4}Pr(Sympt. $\mid$ Plac.) & 0.02 & 0.94 & 0.15 \\
              Pr(Sympt. $\mid$ mAb) & 0.01 & 0.95 & 0.36 \\
              \cmidrule(lr){1-4}Pr(Asympt. $\mid$ Plac.) & -0.01 & 0.94 & 0.17 \\
              Pr(Asympt. $\mid$ mAb) & 0.0 & 0.94 & 0.21 \\
              \cmidrule(lr){1-4}Pr(Sympt. $\mid$ Infec., Plac.) & 0.01 & 0.92 & 0.14 \\
              Pr(Sympt. $\mid$ Infec., mAb) & 0.01 & 0.95 & 0.31 \\
              Pr(Asympt. $\mid$ Infec., Plac.) & -0.02 & 0.92 & 0.16 \\
              Pr(Asympt. $\mid$ Infec., mAb) & 0.0 & 0.95 & 0.14 \\
              \cmidrule(lr){1-4}Pr(N+ $\mid$ Infec., Plac.) & 0.0 & 0.93 & 0.09 \\
              Pr(N+ $\mid$ Infec., mAb) & 0.0 & 0.95 & 0.27 \\
              Pr(N+ $\mid$ sympt., Plac.) & -0.02 & 0.88 & 0.09 \\
              Pr(N+ $\mid$ sympt., mAb) & -0.01 & 0.95 & 0.33 \\
              Pr(N+ $\mid$ asympt., Plac.) & 0.02 & 0.94 & 0.19 \\
              Pr(N+ $\mid$ asympt., mAb) & 0.0 & 0.95 & 0.38 \\
              \cmidrule(lr){1-4}RM TI $\mid$ Plac. & -0.02 & 0.88 & 0.12 \\
              RM TI $\mid$ mAb & -0.01 & 0.92 & 0.06 \\
              \cmidrule(lr){1-4}RM PCR+ $\mid$ Plac. & 0.01 & 0.94 & 0.11 \\
              RM PCR+ $\mid$ mAb & 0.01 & 0.94 & 0.17 \\
              \cmidrule(lr){1-4}Pr(Detected $\mid$ Plac.) & 0.0 & 0.94 & 0.06 \\
              Pr(Detected $\mid$ mAb) & 0.0 & 0.94 & 0.1 \\
              \cmidrule(lr){1-4}PE for infec. & 0.0 & 0.95 & 0.18 \\
              PE for sympt. infec. & 0.0 & 0.96 & 0.16 \\
              PE for asympt. infec. & -0.06 & 0.94 & 0.97 \\
              \cmidrule(lr){1-4}RR for sympt. $\mid$ infec. & -0.01 & 0.95 & 0.34 \\
              RR for N+ $\mid$ infec. & 0.0 & 0.95 & 0.28 \\
              RR for N+ $\mid$ sympt. & 0.01 & 0.96 & 0.35 \\
              RR for N+ $\mid$ asympt. & -0.01 & 0.95 & 0.42 \\
          \bottomrule
    \end{tabular}
    \caption{Bias, coverage, and 95\% confidence interval width of estimates of key quantities obtained by fitting the fully parametric 5 state model shown in \ref{subfig:regendiagram} and \ref{tab:regensim_paraterization} to panel data where PCR and serology were assessed at study day 28. RM TI is the restricted mean time to infection, RM PCR+ is the restricted mean duration of PCR positivity, RR is relative risk, and PE is protective efficacy, calculated as $PE = 1 - RR$.}
    \label{tab:regensim_parametric_res}
\end{table}

\clearpage

\begin{table}[htbp]
    \centering
    \begin{tabular}{rrrr}
          \toprule\multicolumn{4}{c}{Complete results using a semi-parametric 5 state model}\\
          \cmidrule(lr){1-4} & \textbf{Bias} & \textbf{Covg.} & \textbf{CIW} \\
          \cmidrule(lr){2-4}Pr(Infec. $\mid$ Plac.) & 0.0 & 0.94 & 0.07 \\
              Pr(Infec. $\mid$ mAb) & 0.0 & 0.94 & 0.16 \\
              \cmidrule(lr){1-4}Pr(Sympt. $\mid$ Plac.) & 0.01 & 0.95 & 0.16 \\
              Pr(Sympt. $\mid$ mAb) & -0.01 & 0.94 & 0.35 \\
              \cmidrule(lr){1-4}Pr(Asympt. $\mid$ Plac.) & -0.01 & 0.94 & 0.18 \\
              Pr(Asympt. $\mid$ mAb) & 0.0 & 0.94 & 0.21 \\
              \cmidrule(lr){1-4}Pr(Sympt. $\mid$ Infec., Plac.) & 0.01 & 0.94 & 0.14 \\
              Pr(Sympt. $\mid$ Infec., mAb) & -0.01 & 0.94 & 0.31 \\
              Pr(Asympt. $\mid$ Infec., Plac.) & -0.01 & 0.94 & 0.16 \\
              Pr(Asympt. $\mid$ Infec., mAb) & 0.0 & 0.94 & 0.14 \\
              \cmidrule(lr){1-4}Pr(N+ $\mid$ Infec., Plac.) & 0.0 & 0.92 & 0.09 \\
              Pr(N+ $\mid$ Infec., mAb) & 0.0 & 0.95 & 0.27 \\
              Pr(N+ $\mid$ sympt., Plac.) & -0.01 & 0.91 & 0.09 \\
              Pr(N+ $\mid$ sympt., mAb) & 0.01 & 0.96 & 0.33 \\
              Pr(N+ $\mid$ asympt., Plac.) & 0.01 & 0.94 & 0.19 \\
              Pr(N+ $\mid$ asympt., mAb) & -0.01 & 0.94 & 0.39 \\
              \cmidrule(lr){1-4}RM TI $\mid$ Plac. & -0.03 & 0.85 & 0.12 \\
              RM TI $\mid$ mAb & -0.01 & 0.92 & 0.07 \\
              \cmidrule(lr){1-4}RM PCR+ $\mid$ Plac. & 0.02 & 0.88 & 0.11 \\
              RM PCR+ $\mid$ mAb & 0.01 & 0.93 & 0.18 \\
              \cmidrule(lr){1-4}Pr(Detected $\mid$ Plac.) & 0.0 & 0.93 & 0.07 \\
              Pr(Detected $\mid$ mAb) & 0.0 & 0.94 & 0.12 \\
              \cmidrule(lr){1-4}PE for infec. & 0.0 & 0.95 & 0.18 \\
              PE for sympt. infec. & 0.01 & 0.94 & 0.15 \\
              PE for asympt. infec. & -0.06 & 0.94 & 0.96 \\
              \cmidrule(lr){1-4}RR for sympt. $\mid$ infec. & -0.02 & 0.95 & 0.34 \\
              RR for N+ $\mid$ infec. & 0.0 & 0.95 & 0.28 \\
              RR for N+ $\mid$ sympt. & 0.02 & 0.95 & 0.35 \\
              RR for N+ $\mid$ asympt. & -0.01 & 0.95 & 0.42 \\
          \bottomrule
    \end{tabular}
    \caption{Bias, coverage, and 95\% confidence interval width of estimates of key quantities obtained by fitting the semi-parametric 5 state model shown in \ref{subfig:regendiagram} with B-spline baseline intensities (Table \ref{tab:regensim_paraterization}) to panel data where PCR and serology were assessed at study day 28. RM TI is the restricted mean time to infection, RM PCR+ is the restricted mean duration of PCR positivity, RR is relative risk, and PE is protective efficacy, calculated as $PE = 1 - RR$.}
    \label{tab:regensim_semiparametric_res}
\end{table}

\begin{table}[htbp]
    \centering\small
    \begin{tabular}{rrrr}
          \toprule
          \multicolumn{4}{c}{Complete results by crude tabulation of observed data}\\
          \cmidrule(lr){1-4} & \textbf{Bias} & \textbf{Covg.} & \textbf{CIW} \\
          \cmidrule(lr){2-4}Pr(Infec. $\mid$ Plac.) & -0.09 & 0.0 & 0.08 \\
              Pr(Infec. $\mid$ mAb) & -0.2 & 0.0 & 0.16 \\
              \cmidrule(lr){1-4}Pr(Sympt. $\mid$ Plac.) & 0.0 & 0.96 & 0.16 \\
              Pr(Sympt. $\mid$ mAb) & 0.0 & 0.96 & 0.37 \\
              \cmidrule(lr){1-4}Pr(Asympt. $\mid$ Plac.) & -0.19 & 0.02 & 0.17 \\
              Pr(Asympt. $\mid$ mAb) & -0.28 & 0.0 & 0.2 \\
              \cmidrule(lr){1-4}Pr(Sympt. $\mid$ Infec., Plac.) & 0.1 & 0.26 & 0.15 \\
              Pr(Sympt. $\mid$ Infec., mAb) & 0.25 & 0.27 & 0.38 \\
              Pr(Asympt. $\mid$ Infec., Plac.) & -0.11 & 0.26 & 0.17 \\
              Pr(Asympt. $\mid$ Infec., mAb) & -0.11 & 0.27 & 0.17 \\
              \cmidrule(lr){1-4}Pr(N+ $\mid$ Infec., Plac.) & 0.1 & 0.04 & 0.09 \\
              Pr(N+ $\mid$ Infec., mAb) & 0.24 & 0.13 & 0.32 \\
              Pr(N+ $\mid$ sympt., Plac.) & 0.0 & 0.94 & 0.09 \\
              Pr(N+ $\mid$ sympt., mAb) & 0.0 & 0.96 & 0.33 \\
              Pr(N+ $\mid$ asympt., Plac.) & 0.24 & 0.01 & 0.2 \\
              Pr(N+ $\mid$ asympt., mAb) & 0.39 & 0.12 & 0.51 \\
              \cmidrule(lr){1-4}PE for infec. & 0.13 & 0.34 & 0.2 \\
              PE for sympt. infec. & 0.0 & 0.96 & 0.16 \\
              PE for asympt. infec. & 0.39 & 0.73 & 1.07 \\
              \cmidrule(lr){1-4}RR for sympt. $\mid$ infec. & 0.14 & 0.7 & 0.38 \\
              RR for N+ $\mid$ infec. & 0.13 & 0.59 & 0.31 \\
              RR for N+ $\mid$ sympt. & 0.0 & 0.96 & 0.35 \\
              RR for N+ $\mid$ asympt. & 0.13 & 0.81 & 0.46 \\
          \bottomrule
    \end{tabular}
    \caption{Bias, coverage, and 95\% confidence interval width of estimates of key quantities obtained as described in Supplementary Section \ref{subsubsec:crudeests} by crude tabulation of panel data where PCR and serology were assessed at study day 28. RM TI is the restricted mean time to infection, RM PCR+ is the restricted mean duration of PCR positivity, RR is relative risk, and PE is protective efficacy, calculated as $PE = 1 - RR$.}
    \label{tab:regensim_crude_res}
\end{table}

\section{Technical Appendix: Application -- Protective efficacy and natural history in REGEN-2069}
\label{sec:regensupp}

\subsection{Data processing}
\label{subsec:regen_dataprocessing}

\begin{figure}[htbp]
    \centering
    \includegraphics[width=1\linewidth]{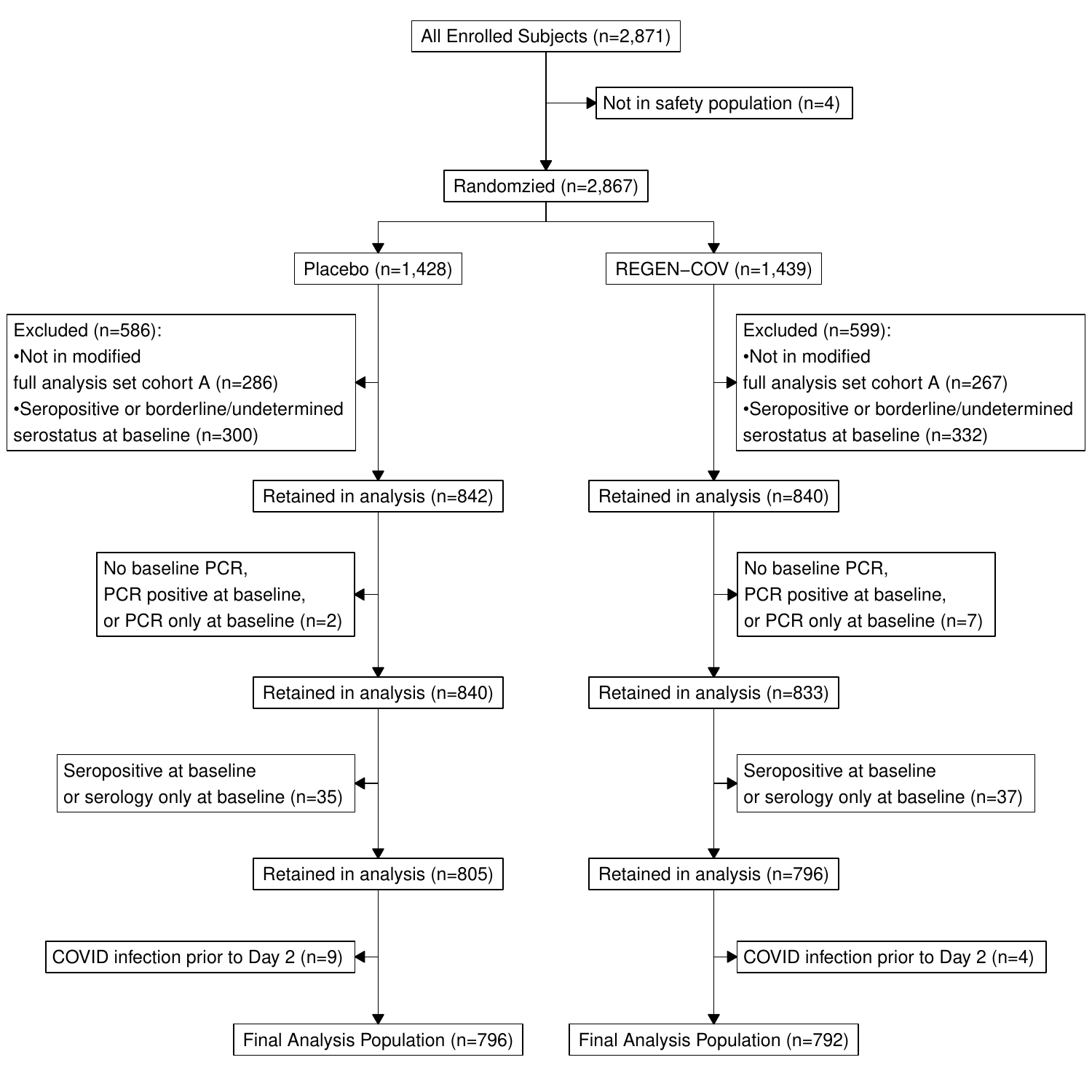}
    \caption{Consort diagram illustrating data processing for REGEN-2069. }
    \label{fig:regen_consort}
\end{figure}

\clearpage
\begin{table}[htbp]
    \centering\small
    \begin{tabular}{ccccc}
    \toprule
    \textbf{Symp.} & \textbf{PCR+} & \textbf{Sero+} & \shortstack{\textbf{mAb} \\ ($N=792$)} & \shortstack{\textbf{Placebo} \\ ($N=796$)} \\
    \cmidrule(lr){1-1}\cmidrule(lr){2-2}\cmidrule(lr){3-3}\cmidrule(lr){4-4}\cmidrule(lr){5-5}
    \xmark & \xmark & \xmark & $746$ & $679$ \\
    \xmark & \xmark & \checkmark & $9$   & $12$  \\
    \xmark & \checkmark & \xmark & $28$  & $22$  \\
    \xmark & \checkmark & \checkmark & $0$   & $27$  \\
    \checkmark & \checkmark & \xmark & $8$   & $13$  \\
    \checkmark & \checkmark & \checkmark & $1$   & $43$  \\
    \bottomrule
    \end{tabular}
    \captionsetup{skip=0.5em}
    \caption{Tabulation of symptomatic infection, positive qRT-PCR, and positive serology in REGEN-2069.}
    \label{tab:regen_summary}  
\end{table}

\clearpage
\subsection{Parameterization of transition intensities}
\label{subsec:regen_model_parameterization}

\begin{table}[htbp]
    \centering
    \def\arraystretch{1.5}
    \begin{tabular}{lll}
    \toprule
        Baseline intensity & Parameterization & Constraints\\
        \cmidrule(lr){1-1} \cmidrule(lr){2-2} \cmidrule(lr){3-3} Exponential& $\log(\lambda_{ij}(t)) = \log(\lambda_{ij}) + \beta Z_i$ & $\lambda_{ij} > 0$\\
        Weibull & $\log(\lambda_{ij}(t)) = \log(\lambda_{ij}) + \log(\alpha) + t(\kappa - 1)  + \beta Z_i$ & $\lambda_{ij} > 0$ and $\kappa_{ij} > 0$\\
        B-spline & $\log(\lambda_{ij}(t)) = \log\left(\sum_{\ell = 1}^L\gamma_\ell B_\ell\left(t;\boldsymbol{\xi},\kappa\right)\right) + \beta Z_i$ & $\gamma_\ell > 0$\\
        \bottomrule
    \end{tabular}
    \caption{Parameterization of transition intensities for the five-state model used in Section \ref{sec:regen2069} and depicted in Figure \ref{subfig:regendiagram}. $Z_i$ is the treatment assignment indicator for participant $i$ with value 0 if $i$ is assigned to placebo and 1 if assigned to mAb. The log baseline intensity from state $i$ to state $j$ is denoted $\log(\lambda_{ij}(t))$.  $B_\ell(t;\boldsymbol{\xi},\kappa)$ is the $\ell-$th basis function for an B-spline with interior knots $\boldsymbol{\xi}$ and degree $\kappa$ evaluated at time $t$.}
    \label{tab:5state_parameterization}
\end{table}

\clearpage
\subsection{Additional results}
\label{subsec:regen_suppresults}

\begin{table}[htbp]
    \centering
    \begin{tabular}{llcc}
      \toprule
      \multicolumn{2}{c}{Parameterization of baseline intensities} & \multicolumn{2}{c}{Results} \\
      \cmidrule(lr){1-2} \cmidrule(lr){3-4}1 → 2 and 2 → 4 & 2 → 3 and 4 → 5 & $\ell(\bY) - \ell_M(\bY)$ & $\Delta$AIC (MCSE) \\
      \cmidrule(lr){1-1} \cmidrule(lr){2-2} \cmidrule(lr){3-3} \cmidrule(lr){4-4}B-Spline($\kappa = 1$; $\xi = [3.5, 10.5]$) & B-Spline($\kappa = 1$; $\xi = 7$) & 100.0 & \textbf{-164.7 (0.19)} \\
      B-Spline($\kappa = 1$; $\xi = [3.5, 10.5, 17.5]$) & B-Spline($\kappa = 1$; $\xi = 7$) & 101.0 & -162.0 (0.19) \\
      B-Spline($\kappa = 1$; $\xi = [3.5, 10.5]$) & B-Spline($\kappa = 0$; $\xi = 7$) & 96.4 & -160.8 (0.19) \\
      B-Spline($\kappa = 1$; $\xi = [3.5, 10.5, 17.5]$) & B-Spline($\kappa = 0$; $\xi = 7$) & 97.5 & -159.0 (0.19) \\
      Weibull & Weibull & 78.7 & -133.3 (0.95) \\
      Exponential & Exponential & 0.0 & 16.0 (0.0) \\
       \bottomrule
    \end{tabular}
    \caption{Model comparison results. The degree and interior knots of the B-Spline intensities are  $\kappa$ and $\xi$, respectively. The quantity $\ell(\bY) - \ell_M(\bY)$ indicates the absolute gain in log-likelihood over the Markov model with exponential transition intensities (larger is better). $\Delta$AIC is computed as $-2(\ell(\bY) - \ell_M(\bY)) + 2p$ (smaller is better), where $p$ is the number of model parameters. The log-likelihood of the Markov model is obtained numerically, while those of the semi-Markov model are estimated via Monte Carlo (Appendix~\ref{subsec:marginal-lik}).}
    \label{tab:regen-aic}
\end{table}

\newpage
\begin{table}[htbp]
    \centering\footnotesize
    \begin{tabular}{ll}
    \toprule
    \textbf{Quantity} & \textbf{Estimate 95\% CI}\\
    \cmidrule(lr){1-1}\cmidrule(lr){2-2}Pr(Infec. $\mid$ Plac.) & 0.146 (95\% CI: 0.121, 0.173) \\
    Pr(Infec. $\mid$ mAb) & 0.058 (95\% CI: 0.042, 0.076) \\
    \cmidrule(lr){1-2}Pr(Sympt. $\mid$ Plac.) & 0.070 (95\% CI: 0.053, 0.089) \\
    Pr(Sympt. $\mid$ mAb) & 0.011 (95\% CI: 0.005, 0.021) \\
    \cmidrule(lr){1-2}Pr(Asympt. $\mid$ Plac.) & 0.077 (95\% CI: 0.058, 0.097) \\
    Pr(Asympt. $\mid$ mAb) & 0.046 (95\% CI: 0.033, 0.063) \\
    \cmidrule(lr){1-2}Pr(Sympt. $\mid$ Infec., Plac.) & 0.475 (95\% CI: 0.381, 0.569) \\
    Pr(Sympt. $\mid$ Infec., mAb) & 0.194 (95\% CI: 0.094, 0.317) \\
    Pr(Asympt. $\mid$ Infec., Plac.) & 0.525 (95\% CI: 0.431, 0.619) \\
    Pr(Asympt. $\mid$ Infec., mAb) & 0.806 (95\% CI: 0.683, 0.906) \\
    \cmidrule(lr){1-2}Pr(N+ $\mid$ Infec., Plac.) & 0.705 (95\% CI: 0.613, 0.791) \\
    Pr(N+ $\mid$ Infec., mAb) & 0.220 (95\% CI: 0.112, 0.358) \\
    Pr(N+ $\mid$ sympt., Plac.) & 0.775 (95\% CI: 0.655, 0.883) \\
    Pr(N+ $\mid$ sympt., mAb) & 0.111 (95\% CI: 0.002, 0.373) \\
    Pr(N+ $\mid$ asympt., Plac.) & 0.643 (95\% CI: 0.522, 0.777) \\
    Pr(N+ $\mid$ asympt., mAb) & 0.246 (95\% CI: 0.122, 0.406) \\
    \cmidrule(lr){1-2}RM TI $\mid$ Plac. & 25.0 (95\% CI: 24.4, 25.5) \\
    RM TI $\mid$ mAb & 26.8 (95\% CI: 26.5, 27.1) \\
    \cmidrule(lr){1-2}RM PCR+ $\mid$ Plac. & 13.0 (95\% CI: 11.5, 14.6) \\
    RM PCR+ $\mid$ mAb & 6.2 (95\% CI: 5.0, 7.8) \\
    Pr(Detected $\mid$ Plac.) & 0.902 (95\% CI: 0.846, 0.941) \\
    Pr(Detected $\mid$ mAb) & 0.690 (95\% CI: 0.582, 0.782) \\
    \cmidrule(lr){1-2}PE for infec. & 0.604 (95\% CI: 0.449, 0.725) \\
    PE for sympt. infec. & 0.836 (95\% CI: 0.694, 0.931) \\
    PE for asympt. infec. & 0.387 (95\% CI: 0.100, 0.608) \\
    \cmidrule(lr){1-2} RR for sympt. $\mid$ infec. & 0.412 (95\% CI: 0.189, 0.677) \\
    RR for N+ $\mid$ infec. & 0.313 (95\% CI: 0.157, 0.502) \\
    RR for N+ $\mid$ sympt. & 0.144 (95\% CI: 0.003, 0.484) \\
    RR for N+ $\mid$ asympt. & 0.386 (95\% CI: 0.185, 0.639) \\\bottomrule
  \end{tabular}
    \caption{Additional results from the analysis of REGEN-2069 data. We abbreviate as follows: Pr = Probability, RMIFT = restricted mean infection free time, RM PCR+ = restricted mean duration of PCR+, PE = protective efficacy, RR = relative risk, N+ = seropositive. RM TI is the arithmetic mean over individuals of the predicted days to infection or maximum follow-up for each individual, whichever is less. RM PCR+ is the arithmetic mean days during which individuals are detectable by PCR or the maximum follow-up for each individual, whichever is less. RR is the ratio of probabilities of the specified event, mAb divided by placebo. PE is 1 - RR, i.e., the reduction in relative risk. }
    \label{tab:regen_fullres}
\end{table}

\newpage
\section{Technical Appendix: Comparison of our proposal distribution for sampling the paths to existing approaches.}
\label{sec:comparison}

\subsection{Comparison with \citet{aralis2019stochastic}}
\label{subsec:araliscomp}

We compare our sampler for the paths to that in \citet{aralis2019stochastic}, which we refer to as the A\&B sampler. The A\&B sampler generates paths from the semi-Markov model via forward simulation and rejects those inconsistent with the observed panel data. Our sampler sample from a surrogate Markov model conditionally on the data. All sample paths are therefore consistent with the data. We then use importance weights to correct for the fact that we sample from a surrogate model.

Table~\ref{tab:comparison-aralis} compares the efficiency of these two samplers for fitting a three-state illness-death model to panel data for 500 individuals. Here, efficiency is measured as the ratio of effective sample size to the number of paths sampled: $ESS/M$. A value of $1$ indicates that the sampler is as efficient as Monte Carlo sampler, while a value of, say, $0.1$ indicates that one needs to sample $10$ times more paths to obtain the same amount of information as a Monte Carlo sampler.

Our sampler significantly outperforms the A\&B sampler. The efficiency of the A\&B sampler decreases as the number of observations per individuals increases, because forward simulation is less likely to generate paths consistent with the observed data, leading to more rejections. In contrast, our approach fares better with richer data, because our sampler conditions on the observed data. Notably, even in the sparser data setting with only 3 observations per individual, our sampler is almost $30$ times more efficient.

The model used for the comparison in Table~\ref{tab:comparison-aralis} has only three states, which favors the A\&B sampler. The performance of this sampler would decrease for models with more states as it becomes less likely to sample a path consistent with the data. In contrast, the efficiency of our sampler is not impacted by the number of states in the model since it conditions on the observed data.

\begin{table}
    \centering
    \begin{tabular}{cccc}
    \toprule
        \shortstack{Number of\\observations\\per individual} & \shortstack{ESS/M for\\our sampler} & \shortstack{ESS/M for\\the A\&B sampler} & Efficiency gain \\
        \midrule
         $3$  & $0.793$ & $0.027$  & $29.8$   \\
         $5$  & $0.833$ & $0.003$  & $308.4$  \\
         $10$ & $0.923$ & $0.0003$ & $3077.5$ \\
         \bottomrule
    \end{tabular}
    \caption{Comparison of the efficiency of our sampler for the paths and that of \citet{aralis2019stochastic} (referred to as the A\&B sampler) in terms of the ratio of the effective sample size (ESS) over the number of sample paths (M).}
    \label{tab:comparison-aralis}
\end{table}

\subsection{Comparison with phase-type Markov models}
\label{subsec:phasetype_comp}

Multistate Markov models with a latent phase-type structure are able to approximate arbitrary time-to-event distributions and have been proposed as tractable approximations for semi-Markov multistate processes \citep{titman2010semi}.  We can fit phase-type directly via numerical optimization, although in some cases (though not for the example below) this may require us to impose constraints on model parameters to ensure that the model is structurally identifiable. These models allow for more flexibility than standard multistate Markov models without a latent phase-type structure. However, there are still known failure modes for phase-type Markov constructions. For instance, such models can have difficulty approximating intensities that are initially close to zero. While this can be ameliorated in some cases by allowing for time-inhomogeneity doing so may create additional computational difficulties due to increasing model complexity. We illustrate this failure mode in a simple case using an illness-death model. We simulate panel data for $1000$ individuals from an illness-death model where the transition healthy$\rightarrow$ill is Weibull with shape $2.5$ and the other transitions are exponential. We consider two models: a Markov multistate model with a phase-type transition healthy$\rightarrow$ill with one latent state, and a semi-Markov model with a spline hazard of degree 1 with 1 knot for the transition healthy$\rightarrow$ill; the remaining transitions are exponential in the two models. We note that the semi-Markov model does not correspond to the model used for simulating the data. The semi-Markov model with a spline hazard offers a much better fit to the data, with an AIC that is 491.2 units lower than the AIC for the Markov model compared with an improvement of 413.9 units of AIC for the phase-type model.

\begin{figure}[htbp]
    \centering
    \includegraphics[width=1\linewidth]{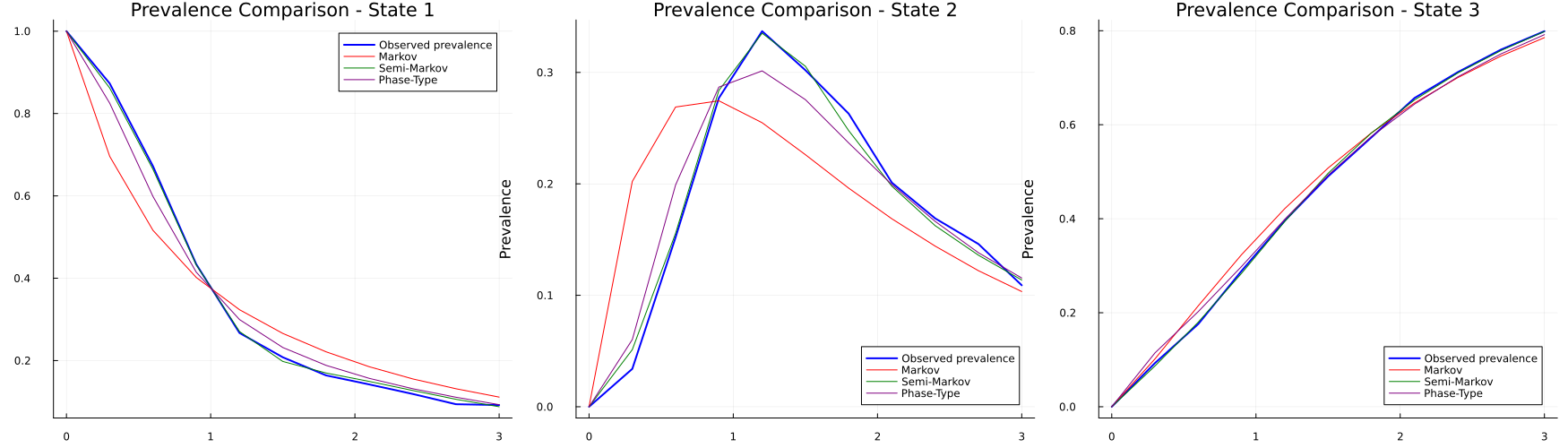}
    \caption{Prevalence curves for the illness-death model under a Markov, a phase-type and a semi-Markov multistate model.
    }
    \label{fig:phase-type-prevalence}
\end{figure}

Figure~\ref{fig:phase-type-prevalence} compares the fitted models and the data with regard to the prevalence curves in each state. The Markov model, with a constant intensity between the healthy and ill states, yields a prevalence in state 2 that increases too early. To compensate for the early increase, the peak of the prevalence in state 2 is much earlier than the observed peak and the rest of the prevalence is too low after time $t=1$. In contrast, the semi-Markov model with a spline intensity closely matches the observed incidence in state 2. Finally, the results of the phase-type model lies between those of the Markov and spline models. The premature increase in the prevalence in state 2  observed for the Markov and phase-type models is due to the transition intensities for the intensities for transitioning from state 1 and eventually into state 2 being too large near time 0. In contrast, under the spline model, the intensity at $t=0$ can be arbitrarily close to $0$.

\begin{table}[htbp]
    \centering
    \begin{tabular}{ccc}
    \toprule
        Model & $\ell(\bY) - \ell_M(\bY)$ & $\Delta$ AIC\\
        \midrule
         Markov  & $0$  & --- \\
         Phase-type -- 2 latent states & $210.9$ & -413.9 \\
         Semi-Markov -- Deg. 1 splines & $\mathbf{247.6}$ & \textbf{-491.2} \\
         \bottomrule
    \end{tabular}
    \caption{Comparison of the goodness of fit of semi-Markov and phase-type models on simulated data from an illness-death model. The quantity $\ell(\bY) - \ell_M(\bY)$ indicates the absolute gain in log-likelihood over the Markov model with exponential transition intensities (larger is better).}
    \label{tab:comparison-phase-type}
\end{table}

\end{document}